\documentclass[11pt]{article}
\usepackage{jheppub}

\usepackage{epsfig}
\usepackage{amssymb}
\usepackage{amsmath}

\usepackage{graphicx, subfigure, array, placeins, float}

\makeatletter \@addtoreset{equation}{section} \makeatother

\graphicspath{{./Artwork/}}

\newcommand{\be}{\begin{equation}}
\newcommand{\ee}{\end{equation}}
\newcommand{\bea}{\begin{eqnarray}}
\newcommand{\eea}{\end{eqnarray}}

\title{Color-kinematic duality for form factors}

\author{Rutger H. Boels,}\emailAdd{Rutger.Boels@desy.de}
\author{Bernd A. Kniehl,}\emailAdd{Bernd.Kniehl@desy.de}
\author{Oleg V. Tarasov,}\emailAdd{Oleg.Tarasov@desy.de}
\author{and Gang Yang}\emailAdd{Gang.Yang@desy.de}
\affiliation{II. Institut f\"ur Theoretische Physik, Universit\"at Hamburg\\ Luruper Chaussee 149, D- 22761 Hamburg, Germany }

\abstract{Recently a powerful duality between color and kinematics has been proposed for integrands of scattering amplitudes in quite general gauge theories. In this paper the duality proposal is extended to the more general class of gauge theory observables formed by form factors. After a discussion of the general setup the existence of the duality is verified in two- and three-loop examples in four dimensional maximally supersymmetric Yang-Mills theory which involve the stress energy tensor multiplet. In these cases the duality reproduces known results in a particularly transparent and uniform way. As a non-trivial application we obtain a very simple form of the integrand of the four-loop two-point (Sudakov) form factor which passes a large set of unitarity cut checks.}

\begin{document}

\maketitle

\setcounter{footnote}{0}


\section{Introduction}

The ability to calculate is of fundamental importance in any would-be theory of nature. In high energy physics this manifests itself in the quest to calculate to higher orders in perturbation theory, including more and more loops and legs. The basic quantities to be calculated in a gauge theory are in general a mixture of plane-wave external states and gauge invariant operators. If only on-shell states are involved these observables are scattering amplitudes, while if only gauge invariant operators are involved these are pure correlation functions. Admixtures of observables build of operators and on-shell states are called form factors. 

The main motivation of this paper is the observation that in recent years many new insights have been developed for the calculation of scattering amplitudes in gauge theories (see e.g. \cite{Dixon:2011xs} and references therein), while similar developments for form factors and correlation functions have been minor in comparison. This paper is part of an effort to redress this imbalance. Recent developments have shown that maximally supersymmetric Yang-Mills theory is a good testing ground for new ideas which then should trickle down to non-supersymmetric gauge theory. This line of thought will also be followed here. 

A particular insight that has recently been developed is a duality between color and kinematics that has been proposed for integrands of scattering amplitudes \cite{Bern:2008qj} \cite{Bern:2010ue}. This duality will be extended here to form factors. Color-kinematic duality in its current form states that at tree and loop level the integrand of a scattering amplitude can be arranged in such a way that whenever the color factors of an integrand obey a Jacobi relation, the kinematic parts of these diagrams can also be chosen such that they obey the same identity. In a second step which will not be of much importance for the purposes of this article, this duality can also be used to construct the integrand of a scattering amplitude in a gravity theory, whose field content is simply a tensor product of two gauge theory copies. The status of color-kinematic duality will be reviewed below. 

There are two distinct ways the ideas of color-kinematic duality could apply to more general observables which include gauge invariant operators. The first is simply as a direct consequence of color-kinematic duality for scattering amplitudes. On a unitarity cut and at tree level poles of quite general form factors the duality should apply to the amplitude parts of the residues and discontinuities, see for instance \cite{Engelund:2012re}. In this paper we conjecture that the duality can be used directly for general form factors and in particular without taking any cuts. We will show in examples that the duality can be used as a powerful tool to produce consistent ansatz for high loop form factors. In particular, our results unify and greatly simplify the results in the recent literature up to three loops. 

As a new non-trivial application, we compute for the first time the four-loop integrand of the Sudakov form factor in ${\cal N}$=4 super Yang-Mills theory (SYM) in a form which passes all (highly non-trivial) unitarity checks performed. The existence of such a form for the four-loop form factor is a strong evidence that the color-kinematic duality holds in general. A distinct advantage of using color-kinematic duality is that the traditionally hard to obtain non-planar integrals in the problem are fixed by the coefficients of planar integrals. The evaluation of the found integrals is in progress \cite{inprogress}. 

This article is structured as follows. In section \ref{sec:review} the necessary background for this paper will be reviewed. This is followed by section \ref{sec:gensetup} where our general calculational strategy is explained. This strategy is applied to a few examples in section \ref{sec:examples}, and in particular the four-loop form factor in section \ref{sec:fourloop}. The main presentation ends with a discussion and conclusion section. In the first appendix we give a representative selection of four-loop Jacobi relations. The other appendix briefly discusses color structures which appear in two-point form factors and correlation functions with two adjoint operators through eight loops.


\section{Review}\label{sec:review}


\subsection{Color-kinematic duality for scattering amplitudes}

Color-kinematic duality is the collective name for what is basically a pair of two partially proven conjectures \cite{Bern:2008qj}, \cite{Bern:2010ue}. The starting point is to write the integrand for a scattering amplitude in $D$ dimensions at $l$ loops in terms of a sum over trivalent graphs only,
\begin{equation}
A^{(l)} = \sum_{\Gamma_i} \int \prod_{j=1}^l d^D \ell_j  \frac{1}{S_i} \frac{n_i c_i}{s_i} ,
\end{equation} 
where $c_i$ is the natural color factor associated to trivalent graph $i$, $S_i$ is the symmetry factor associated to the trivalent graph $i$ and $s_i$ is the product of all internal propagators of the graph. This form of the integrand simply amounts to a rewriting of Feynman graph-based perturbation theory by parcelling out the four-vertex over three-vertices. This determines some set of numerators $n_i$ which is not unique. The non-trivial claim of color-kinematic duality is that there exists a set of numerators such that whenever color factors obey a Jacobi identity, the numerators do too:
\begin{equation}
\forall \{i,j,k\} \quad c_i + c_j + c_k = 0 \quad \Rightarrow\quad  n_i + n_j + n_k = 0  \, .
\end{equation}  
A set of numerators which obeys all Jacobi identities is called color-dual. Color-dual numerators have been constructed explicitly at tree level for all multiplicities through a variety of methods \cite{Kierm:2010xq,Mafra:2011kj,BjerrumBohr:2012mg}, and at loop level in specific examples in ${\cal N}=4$ SYM for five points up to two loops \cite{Carrasco:2011mn} and for four points up to four loops \cite{Bern:2012uf}. Moreover, in pure Yang-Mills theory numerators are known at two loops  for the four-point helicity equal amplitude \cite{Bern:2010ue} and at one loop for the helicity equal and one-unequal cases \cite{Boels:2013bi}.  Of course, the tree level results will extend almost trivially to any generalized cut of a loop amplitude which involves tree amplitudes. 

The second conjecture usually taken to be part of color-kinematic duality is that if a set of color-dual numerators exists in a gauge theory, then an amplitude in a gravitational theory may be constructed as
 \begin{equation}
M^{(l)} = \sum_{\Gamma_i} \int \prod_{j=1}^l d^D \ell_j  \frac{1}{S_i} \frac{n_i \tilde{n}_i}{s_i} \, ,
\end{equation} 
where $n$ and $\tilde{n}$ are color-dual numerators for two in general distinct gauge theories. This conjecture has been proven at tree level, assuming that local numerators exist \cite{Bern:2010yg}. The field content of the gravity theory is the direct product of the field contents of the gauge theories. In this way 'squaring' $\mathcal{N}=4$ SYM gives the maximal (ungauged) $\mathcal{N}=8$ super-gravity theory. Similarly, squaring pure Yang-Mills theory gives Einstein gravity coupled to a dilaton and a two-form. The latter theory is sometimes referred to as $\mathcal{N}=0$ supergravity. The  color-kinematic duality relation of gauge theories to gravitational theories will not play a central role in this article. 

As an illustration of the general idea, consider the four-point tree amplitude in gauge theory which can be written in terms of three trivalent diagrams as
\be {\cal A}^{(0)}_4(1, 2, 3, 4) \, = \, {c_s \, n_s \over s} + {c_t\, n_t \over t} + {c_u \, n_u \over u} \ . \label{eq:BCJfourpoint} \ee
%
%
\begin{center}
\begin{figure}[h]
\centerline{\includegraphics[height=2.3cm]{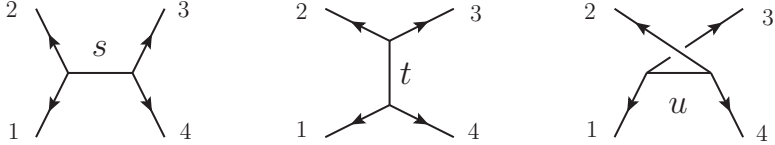} } \caption{\it
Trivalent diagrams for four-point tree amplitudes.} \label{treeA4}
\end{figure}
\end{center}
%
The color factors denoted as $c_i$ are given by the product of $\tilde f^{abc}$ associated to each trivalent vertex
\be c_s =\tilde f^{a_1 a_2 b}\tilde f^{b a_3 a_4}, \qquad c_t =\tilde f^{a_2
a_3 b}\tilde f^{b a_4 a_1}, \qquad c_u =\tilde f^{a_1 a_3 b}\tilde f^{b a_2 a_4} \, ,
\ee
where
\be \tilde f^{abc} = i \sqrt{2} f^{abc} = {\rm Tr}( [T^a, T^b] T^c) \, . \label{tildef}  \ee
They satisfy the Jacobi identity
\be c_s = c_t + c_u  \ . \ee
By color-kinematic duality one can also have %
\footnote{Note that we have necessarily made some choice of signs for the color factors. One can choose other conventions, and the relation between $n_i$ should change correspondingly. The final result is given in terms of the product $c_i\, n_i$, and therefore is independent of this choice. }
\be n_s = n_t + n_u  \ . \ee
The solution to the Jacobi conditions may be given explicitly as
\be \label{nstu} n_s = \alpha \ , \qquad n_t = - t \, A^{(0)}_4(1,2,3,4) - \alpha \, {t \over s} \ , \qquad n_u = t \, A^{(0)}_4(1,2,3,4) - \alpha \, { u \over s} \ . \ee
This can be shown to reproduce the full color-dressed amplitude. The solution depends on an arbitrary parameter $\alpha$. This freedom in specifying numerators is referred to as a generalized gauge transformation  \cite{Bern:2008qj}. In general this gauge freedom is the freedom to choose another set of kinematic numerators
\be n'_i = n_i + \Delta_i \ , \ee
%
%
\begin{center}
\begin{figure}[t]
\centerline{\includegraphics[height=2.7cm]{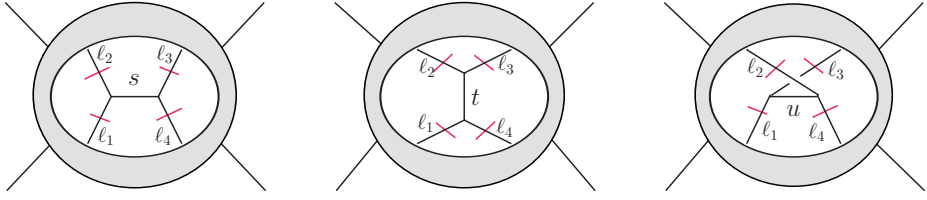} } 
\caption{\it
Color-kinematics duality at loop level.} \label{loopBCJ}
\end{figure}
\end{center}
%
which do not change the tree amplitudes as long as 
\begin{equation}
\sum_{\Gamma_i} \frac{\Delta_i c_i}{s_i} = 0
\end{equation}  
holds. Moreover, the $\Delta$ should satisfy Jacobi identities.

It should be obvious that through unitarity cuts tree-level color-kinematics duality provides constraints to the loop integrand. This occurs whenever a tree amplitude is isolated by the cuts. For example as shown in Figure \ref{loopBCJ}, if we cut four propagators $\ell_1, .., \ell_4$, we have%
\footnote{Sums over all states to appear in the cuts are understood but not shown explicitly in the formula. }
\bea && {\cal A}_n^{(l)}
\Big|_{(\ell_1^2, .., \ell^2_4)-{\rm cut}} 
= \int\!\!d{\rm LIPS} (\ell_1, .., \ell_4) \
{\cal A}_{n+4}^{(l-4)} \, {\cal A}^{(0)}_4(\ell_1, \ell_2, \ell_3, \ell_4)
\nonumber\\ &=&  \int\!\!d{\rm LIPS}  (\ell_1, .., \ell_4) \
\left( \int\prod_i^{l-4} d^D \ell_i \sum_j {c_j n_j \over \prod_a D_a} \right) \, \left[ {c_s n_s \over (\ell_1 +
\ell_2)^2} +  {c_t n_t \over (\ell_1 + \ell_3)^2} +  {c_u n_u
\over (\ell_1 + \ell_3)^2} \right] 
\nonumber\\ &=&  \int\!\!d{\rm LIPS}  (\ell_1, .., \ell_4) \
\int \prod_i^{l-4} d^D \ell_i \sum_j {1 \over \prod_a D_a } \left[ {C_{j,s} N_{j,s} \over (\ell_1 +
\ell_2)^2} +  {C_{j,t} N_{j,t} \over (\ell_1 +
\ell_4)^2} +  {C_{j,u} N_{j,u} \over (\ell_1 +
\ell_3)^2} \right] \, ,  \eea
where in the second line the Bern-Carrasco-Johansson (BCJ) \cite{Bern:2008qj} representation of four-point tree amplitudes of equation \eqref{eq:BCJfourpoint} was used and ${\cal A}_{n+4}^{(l-4)}$ has been expressed in terms of some loop integrals. The third line is simply given by defining
\be C_{j,s} \equiv c_j \, c_s , \qquad N_{j,s} \equiv n_j \, n_s \ . \ee
Because of $ n_s = n_t + n_u$, we have
\be   N_{j,s} \big|_{\ell^2_1= \ell_2^2 = \ell_3^2 =\ell^2_4=0} = (N_{j,t} + N_{j,u}) \big|_{\ell^2_1= \ell_2^2 = \ell_3^2 =\ell^2_4=0}  \, , \ee
which may be thought of as the dual relation of that of color factors, $C_{j,s}  = C_{j,t} + C_{j,u}$. Beyond amplitudes it was noted in \cite{Engelund:2012re} that a similar observation should hold for unitarity cuts\footnote{This includes the case of tree level poles.} of general form factors, whenever a scattering amplitude is isolated on the cuts.  

Beyond the constraints on loop integrands from unitarity cuts it was conjectured that there exists a representation where the above duality relations are truly  {\it off-shell}, i.e. without doing cuts to impose on-shell conditions \cite{Bern:2010ue}. More explicitly, given a representation that
\be  {\cal A}_n^{(l)} = \sum_{\Gamma_i} \int \prod_j^l d^D \ell_j {1\over S_i} {C_i \, N_i \over
\prod_a D_a} \, , \ee
there is an off-shell color-kinematics duality
\be C_i = C_j + C_k \quad \Rightarrow \quad N_i = N_j + N_k \, . \ee
This is a highly non-trivial claim. This has been checked on a case-by-case basis, but no general proof exists. Some supporting evidence was recently uncovered in \cite{Oxburgh:2012zr}. It may be taken as an evidence of the existence of a Lagrangian formulation of gauge theory which has explicitly color-dual structure. This much more powerful version of color-kinematic duality at loop level will be generalized to a class of form factors in this article.


\subsection{Form factors in $\mathcal{N}$=4 supersymmetric Yang-Mills theory}

As recalled in the introduction, form factors are the most generic class of observables in any Yang-Mills theory. They are mixtures of $i$ gauge invariant operators $O(x)$ and $j$ on-shell states $s$,
\begin{equation}
\tilde{F}(O_1 \ldots O_i, s_1, \ldots s_j)
\end{equation}
In this article only the simplest form factors with one insertion of a gauge invariant operator will be considered. This will be referred to as a $j$-point form factor. The simplest example is the two-point form factor which is also called the Sudakov form factor. In this article these form factors are considered in the context of $\mathcal{N}=4$ SYM. The operator inserted will belong to the half-BPS stress-energy multiplet: the multiplet containing the (on-shell) stress-energy tensor as well as the self-dual and anti-self-dual field strengths. Another member of the multiplet which is usually useful for practical calculation is 
\begin{equation}
O = \textrm{Tr} (\phi^{34} \phi^{34}) \, ,
\end{equation}
where the field $\phi^{IJ}$ with anti-symmetric $SO(6)$ indices parameterizes the six scalars of the  $\mathcal{N}=4$ super Yang-Mills theory. 

The object of study in this article is the form factor with the operator Fourier transformed to momentum space, i.e.
\be F(1, \ldots , n; q ) \, = \,  \int\, d^4x \, e^{-iqx} \, \langle 1 \cdots  n | {\cal O} (x) |0\rangle \ . \ee
The momentum of the inserted operator in general does not square to zero, $q^2 \neq 0$. In this sense it is 'off-shell'. This does not mean it does not satisfy field equations: the conservation of the stress-energy tensor in field theory for instance is typically only guaranteed up to field equations. For the form factor these are satisfied for all inserted operators in the problem. 

Compared to progress on amplitude computation, form factors are much less studied and have only attracted renewed attention very recently, see for example \cite{vanNeerven:1985ja, Alday:2007he, Maldacena:2010kp, Brandhuber:2010ad, Bork:2010wf, Brandhuber:2011tv, Bork:2011cj, Henn:2011by, Gehrmann:2011xn, Brandhuber:2012vm, Bork:2012tt}. The current list of achievements at weak coupling beyond one loop in $\mathcal{N}=4$ SYM is 
\begin{itemize}
\item 2pt @2 loops: \cite{vanNeerven:1985ja},
\item 2pt @3 loops: \cite{Gehrmann:2011xn},
\item 3pt @2 loops: \cite{Brandhuber:2012vm}.
\end{itemize}
Moreover, the three-loop result was only obtained \emph{after} the much harder QCD computation \cite{Baikov:2009bg} (see also \cite{Gehrmann:2010ue}). The same holds for  the two-loop three-point computation whose QCD counterpart can be found in \cite{Gehrmann:2011aa}. Form factors at strong coupling (in 2D kinematics) were considered in \cite{Alday:2007he, Maldacena:2010kp}. In the following mainly two- and three-point form factors will be considered, unless indicated otherwise. These have both special kinematics as well as special color properties. The two-point form factor has to be proportional to the color delta function and can only be a function of the momentum squared of the operator leg $q^2$ as well as, in $SU(N_c)$ gauge theory, the number of colors $N_c$ and the gauge coupling $g_{\textrm{ym}}$. Moreover, the two-point form factor has to be proportional to the tree form factor by on-shell supersymmetry,
\begin{equation}\label{eq:genformfac2pts}
F^{(l)}(1,2; q) = \delta^{ab} F^{(0)}(1,2,q) f(q^2, N_c, g_{\textrm{ym}}) \, .
\end{equation}
In particular, the function $f$ does not depend on the inserted operator. As a function of $N_c$ and $g_{\textrm{ym}}$ it can be rewritten in terms of powers of the 't Hooft coupling, 
\be \lambda  = g^2_{\textrm{ym}} N_c \, , \ee
and, in general, at a finite loop order as a finite taylor series in $N_c^{-1}$. Non-trivial terms in this series are called non-planar corrections. Actually, as far as the color structure is concerned, the two-point form factor is the same as the two-point correlator which is computed to extract the $\beta$ function at high loop order \cite{vanRitbergen:1997va}. From this it follows that the first non-planar correction to the two-point form factor occurs at four loops, regardless of gauge theory under study. This point is worked out more fully in appendix \ref{app:color}. 

For general gauge group the number of colors $N_c$ in an $SU(N_c)$ calculation  is replaced by appropriate Casimir operators. The three-point form factor can have more interesting dynamics and more complicated color factors. In particular, in addition to the structure constant for the gauge group the completely symmetric structure constant $d^{abc}$ might make an appearance. On-shell supersymmetry still restricts the form factor to be proportional to the tree. Since the latter is known to be absent at two loops \cite{Brandhuber:2012vm} in $\mathcal{N}=4$ SYM, the first appearance of this can be at three loops. Based on the details of the computation of the beta function at three loops in \cite{Avdeev:1980bh,Tarasov:1980au, Tarasov:1980kx} it seems likely it indeed does appear at this loop order. For more points more general color structures and operator-dependent pre-factors can and do appear.


\section{General formalism \label{sec:gensetup}}

In this section first color-kinematic duality will be generalized to form factors. This is first done at tree level for clarity and then extended to loop level. In the second half of the section we outline how this conjecture can be used in general to compute the integrand of the form factor. This technique is then applied in the following sections in specific examples. 

\subsection{Color-kinematic duality for form factors}
The color structure of form factors is different from that of amplitudes. To get our bearings a few explicit examples will be studied first. The simplest example is the two-point case which always has the trivial color factor $\delta^{ab}$ where $a$ and $b$ are the color indices of the gluons. In particular, there is no Jacobi identity. This also holds for the three-point form factor at tree level as it can be written as the sum of three trivalent diagrams as 
\be {\cal F}_3^{(0)} = {\tilde f^{a_1a_2a_3} \, \delta^{(8)}(\lambda_1 \eta_1 + \lambda_2 \eta_2 + \lambda_3 \eta_3) \over \langle12\rangle \langle 23\rangle \langle 31\rangle} = {\tilde f^{a_1a_2 b} \, \delta^{a_3b} \, n_{12} \over s_{12} } + {\tilde f^{a_2 a_3 b} \, \delta^{a_1 b} \, n_{23} \over s_{23} } + {\tilde f^{a_3a_1b} \, \delta^{a_2 b} \, n_{31} \over s_{31} } \, , \ee
where 
\be n_{ij} = {s_{ij} \over 3\langle12\rangle \langle 23\rangle \langle 31\rangle}\delta^{(8)}(\lambda_1 \eta_1 + \lambda_2 \eta_2 + \lambda_3 \eta_3)  \, . \ee
The fermionic delta function is included, see e.g. \cite{Brandhuber:2010ad, Bork:2010wf}. Hence this form factor is proportional to a single structure constant. 

\begin{center}
\begin{figure}[h]
\centerline{\includegraphics[height=3cm]{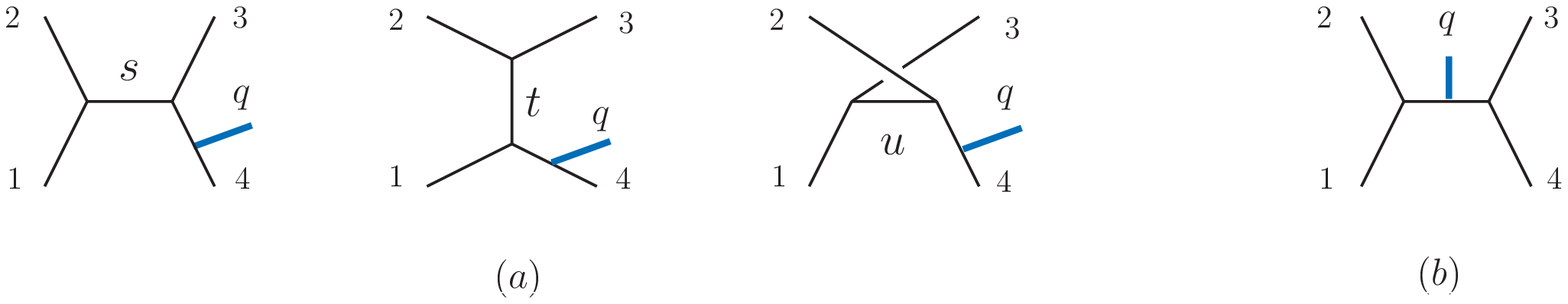} } \caption{\it
Diagrams for four-point tree form factor.} 
\label{treeF4}
\end{figure}
\end{center}

The four-point case becomes more non-trivial. There are $15$ different trivalent diagrams: there are five ways the gauge invariant operator can be inserted in the trivalent graphs of a four-particle amplitude. Some of the graphs are illustrated in figure \ref{treeF4}. On poles of the Feynman graphs where a tree level amplitude is isolated color kinematic duality should hold. Consider for instance the sum of the three diagrams shown in Figure\,\ref{treeF4}(a),
\be \label{D4} D^{(4)} = {\delta^{a_4 c} \over s_{123} } \left( {\tilde f^{a_1a_2 b}\tilde f^{ba_3c} \, n_{12}^{(4)} \over s_{12} } + {\tilde f^{a_2a_3b}\tilde f^{bca_1} \, n_{23}^{(4)} \over s_{23} } + {\tilde f^{a_1a_3b} \tilde f^{ba_2c} \, n_{13}^{(4)} \over s_{13} } \right) \, . \ee
%
%
%
At the kinematic pole $s_{123}=0$ the residue given by the expression in the bracket becomes a four-point amplitude. Color-kinematic duality for the tree amplitude then imposes
\be n_{12}^{(4)} \big|_{s_{123}=0} =  \left( n_{23}^{(4)} + n_{31}^{(4)} \right) \big|_{s_{123}=0} \, . \ee
This is the lesser sense in which color kinematic duality should hold for form factors. The conjecture we make in this paper is that the duality also holds before taking any poles, in a sense to be made precise below.

Before writing more general formulas, let us illustrate the idea in the case of the tree level four-point form factor. In this case the form factor can be written over two classes of diagrams: those with the operator inserted on an external leg (class $D$, see equation \eqref{D4}) or the class with the operator inserted on the internal leg ($E$),
\be {\cal F}_4^{(0)} = \sum_{i=1}^4 D^{(i)}  + E \label{eq:genconj} \, . \ee
Explicitly, $E$ follows from the diagrams such as shown in Figure\,\ref{treeF4}(b) 
 \be E = { \delta^{bc}\tilde f^{a_1 a_2 b}\tilde f^{a_3 a_4 c} \,n_{(12,34)}  \over s_{12}\, s_{34} } + { \delta^{bc}\tilde f^{a_2 a_3 b}\tilde f^{a_4 a_1 c} \,n_{(23,41)}  \over s_{23}\, s_{41} }+  { \delta^{bc}\tilde f^{a_1 a_3 b}\tilde f^{a_2 a_4 c} \,n_{(13,24)}  \over s_{13}\, s_{24} } \, , \ee
We conjecture that numerators can always be found such that they mirror the Jacobi identities of the color structures they multiply, i.e.
\be n_{12}^{(i)} =  n_{23}^{(i)} + n_{31}^{(i)} \, , \ee
as well as
\be n_{(12,34)} =  n_{(23,41)} + n_{(13,24)} \, . \ee

This is quite plausible as a simple counting argument shows. At tree level, there is a minimal basis for color factors modulo all Jacobi relations as explained in \cite{DelDuca:1999rs} (see also \cite{Boels:2012sy}). For four particle color factors (``colormatics'') this minimal solution has two elements. Implementing this for both color as well as kinematic factors in equation \eqref{eq:genconj} gives two equations in terms of $10$ unknowns, i.e. 2 for each $D$ and 2 for the $E$ type factors. This set of equations has many solutions. In fact, it is straightforward to check that the resulting equations have solutions with additional symmetries. In particular, one would expect that the different $D$ type diagrams are related by simple interchange symmetries.  

One solution is to set all $E$ type numerators to zero and
\begin{eqnarray}
 n^{(4)}_{12} &=& \frac{1}{4}  \left(s_{12} (s_{23} + s_{24})  F(1234) + s_{12} s_{24} F(1324)  \right) \, ,\\
n^{(4)}_{23} &=& \frac{1}{4}\left(s_{24} (s_{23} + s_{12})  F(1324) + s_{12} s_{24} F(1234)  \right) \, ,
\end{eqnarray}
where 
\be s_{ij} = (\hat{p}_i + \hat{p}_j)^2 \quad  \textrm{with}  \quad \begin{array}{rl} \hat{p}_i & = p_i \, , \qquad i=1,2,3 \\ \hat{p}_4 & = p_4 + q \, , \end{array}  \ee
and $F(1234)$ as well as $F(1324)$ are color-ordered form factors with the indicated ordering. The other $n^{(i)}$ are simply defined by interchanging $4$ with $i$ in the above formula. This solution was obtained by explicitly solving the system of equations. 

In particular, in this solution it is seen that no constraints on the form factors are necessary: there are no BCJ-type  \cite{Bern:2008qj} relations for the four-point tree level form factor. This is to be contrasted with the four-point amplitude computation where a similar computation shows that the duality imposes additional relations on the amplitudes. The color kinematic duality is expected to only hold for amplitude subgraphs separately.  The sets of color-dual numerators do not in general sum to anything physical separately.

The above observation for four-point form factors can be generalized to form factors at both tree and loop level. At tree level, one should decompose the tree level form factor in terms of the place the gauge invariant operator is inserted,
\be {\cal F}_n^{(0)} = \sum_{i} {\cal{F}}_n^{(i),0} \, .\ee
Then for each ${\cal{F}}_n^{(i),0}$ separately we conjecture color-dual numerators can be found. There is a considerable number of  freedom to choose the numerators as a simple counting argument of equations and unknowns will yield.   

\subsection*{Color-kinematic duality for loop form factors}
At loop level a similar decomposition may be imposed. Then for every ${\cal{F}}^{(i),l}$ one introduces numerators and color factors as in the amplitude integrand case,
\begin{equation}
 {\cal{F}}_n^{(k),l}  = \sum_{\Gamma_i} \int \prod_{j=1}^l dl^D_j  \frac{1}{S_i} \frac{n^{(k)}_i c_i}{s^{(k)}_i} \, .
\end{equation} 
Moreover, one can impose numerator symmetries as in the $D$-type contributions above: if two diagrams are related by a symmetry, then their numerators are related. At tree level, by an extension of the counting argument above it is quite plausible that color-dual numerators always exist. At loop level it will be shown below that the conjecture gives strong constraints for the integrands of form factors. 

This conjecture at loop level will be explored below in several examples in the next section. In the remainder of this section a complete overview is given over the calculational setup. The basic outline of this mirrors to quite some extent the setup for scattering amplitudes up to four loops as described in \cite{Bern:2012uf}.

\subsection{Calculational strategy}

The full integrand of a $L$-loop form factor can be given as 
\be {\cal F}_n^{(L)} \, = K_n \, \sum_{\sigma_n} \sum_i {1\over S_i} \, C_i \, I_i  \ , \ee
where we sum over a set of basis integrals,%
\footnote{The coupling constant $g_{ym}$ will usually be suppressed in our results.}
\be I_i = (-i)^L \int \prod_{j=1}^L {d^D \ell_j \over (2\pi)^D} {N_i(p_j,\ell_m) \over \prod_a D_a} \ ,  \ee
which are associated to $L$-loop diagrams with only trivalent vertices. The classification of trivalent graphs will be  discussed in the following subsection. We sum over all permutations  $\sigma_n$ of external on-shell legs. $K_n$ is some uniform kinematic factor. For two-point and three-point cases  that we will consider
\be K_2 = s_{12}^2 {\cal F}_2^{(0)} \, , \qquad K_3 = {\cal F}_3^{(0)} \, . \ee

The $C_i$'s are color factors. For a given trivalent graph, each cubic vertex can be dressed with a factor $\tilde f^{abc}$.
In the form factor there is also a special vertex, the one connected to the off-shell leg $q$ that is dressed with the factor $\delta_{ab} = {\rm Tr}(T^a T^b)$. The whole color factor $C_i$ is then given by taking the product of all $\tilde f^{abc}$'s and the single $\delta_{ab}$. The $S_i$'s are symmetry factors. They account for the over-counting both from the sum over the permutations of external on-shell legs and the internal symmetries of the graph. In practice, $S_i$ for a given graph can be  computed by randomly shuffling the list of edges and then counting the number of inequivalent isomorphic maps.

The main challenge is to determine the numerator $N_i$ for each basis integral. In principle, one can apply Feynman rules and sum over all Feynman diagrams. However, this becomes very complicated for higher loops, in particular for ${\cal N}$=4 SYM with its large number of fields. An alternative is to construct the amplitude based on  traditional unitarity methods.

Here the conjectured color-kinematics duality for the loop integrand is used to construct a consistent ansatz which is then verified through unitarity cuts. For a given trivalent graph, one can apply the Jacobi relations to relate it to two other graphs, such as those shown in Figure\,\ref{loopBCJ},
\be C_i = C_j + C_k \quad \Rightarrow \quad N_i = N_j + N_k \, . \ee
Note that no cuts for loop propagators are imposed here. For each propagator (except the two connected to the off-shell leg $q$), there is one such equation. By considering all graphs, one can obtain a set of equations for the numerators. This gives strong constraints on the numerators. The main job is to find a solution which satisfies the full set of equations, while also consistent with the unitarity cuts.  Below the various substeps in the calculation are discussed in yet more detail.


\subsubsection{Generating topologies}

The first step in applying color-kinematics duality is to generate all trivalent graphs at a given loop order. There are mature techniques to generate inequivalent topologies; we have opted for \cite{DiaGen} because of its simplicity and ability to handle higher loop orders. Manipulations of graphs have generically been performed with Mathematica. For the problem at hand we need to make some particular choices. 

\begin{enumerate}

\item The diagrams are trivalent, one-particle irreducible graphs, without bubble sub-graphs. For the form factor graph the vertex which is connected to the off-shell leg $q$ (corresponding to the operator) is special.

\item Since we mainly focus on ${\cal N}$=4 SYM, we further exclude most diagrams which contain triangle subgraphs, due to the no-triangle property of ${\cal N}$=4 SYM \cite{Bern:1994zx}. For form factors,  a single triangle is allowed if it is connected to the off-shell leg $q$.  The triangle subgraphs we exclude are the one-loop and two-loop non-planar triangles%
\footnote{
For two-loop non-planar sub-triangles this concerns graphs like 
\begin{tabular}{c}{\includegraphics[height=1.3cm]{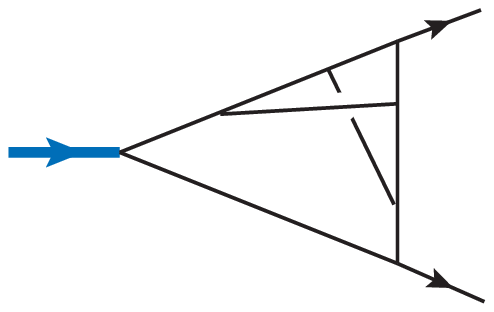}}\end{tabular}.
}. 
This is enough up to four loops. For five and higher loops, there can be three-loop or higher-loop triangle subgraphs, which may be excluded and can reduce further the number of topologies.  (This further subtraction is not considered in the following table.)

\item The diagrams usually have the maximal number of loop propagators. In the construction with color-kinematics duality, one also needs to consider the graphs with less loop propagators but with more complicated trivalent tree legs, for example, the ``snail-like" diagrams \cite{Bern:2012uf}. Such diagrams can be easily obtained by replacing a single leg with multi-leg trivalent tree graphs in lower-point cases.  One could also generate these graphs by applying color-kinematics relations to those graphs containing maximal numbers of propagators. (In the following table only graphs with maximal numbers of loop propagators are counted.)

\end{enumerate}

Based on the above selection rules, we obtain the number of different topologies for two- and three-point form factors and four- and five-point amplitudes up to six loops, as shown in Table\,\ref{tab-numtop}.

\begin{table}[ht]
\begin{center}
\caption{\it The number of topologies of trivalent graphs allowed by the selection rules.
\label{tab-numtop}
}
\vskip .5 cm
\begin{tabular}{l | c | c | c | c | c | c } 
\# of loops & 1  &  2 & 3 & 4 & 5 & 6   \cr \hline \hline
\# of topologies for $A_4$ $\begin{matrix} ~ \\ ~   \end{matrix}$   & 1 & 2 & 9 & 52 & 446 & 4891  \cr \hline
\# of topologies for $F_2$ \begin{tabular}{cc}  ~ \\  ~ \end{tabular}   &  1 & 2 & 6 & 34 & 273 & 2718  \cr \hline \hline
\# of topologies for $A_5$ $\begin{matrix} ~ \\ ~   \end{matrix}$   & 1 & 3 & 19 & 155 & 1684 & 22225  \cr \hline
\# of topologies for $F_3$ \begin{tabular}{cc}  ~ \\  ~ \end{tabular}   &  1 & 4 & 22 & 171 & 1695 & 20046  \cr \hline
\end{tabular} 
\end{center}
\end{table}

From the table it follows that for higher-loop two-point form factors the number of graphs grows about as fast as for the four-point amplitude, but is typically smaller than the four point amplitude by a factor of $2$.


\subsubsection{Solving color-kinematics dualities}

After generating the trivalent graphs, the constraints imposed by color-kinematic duality can be studied. 

\begin{enumerate}

\item Given a trivalent graph, for each internal propagator (except the two connected to the off-shell leg $q$), one can apply the Jacobi identity to generate two other graphs. These two graphs can both be part of the basis set. Often, one of the two contains a triangle subgraph. In this case the corresponding numerator is set to zero. By considering all propagators of all graphs, one generates a set of equations for the numerator functions of the trivalent graphs in the basis.

\item Typically the numerators of a small set of graphs can be identified from which all other numerators can be obtained by the set of equations derived from the duality. The corresponding graphs for these numerators are called \textit{master integrals} since \cite{Carrasco:2011mn}. These can be identified by inspecting the full set of equations: there is usually no unique choice.

\item One can construct the most general ansatz for the master integrals, under several simple constraints derived from physical expectations:
	\begin{itemize}
	\item One would expect the numerators to be local. In particular they should be polynomial functions of momenta (both loop and external).
	\item The full expression should not break the excellent UV properties of $\mathcal{N}=4$ SYM. In practice this implies that for any $n$-point one-loop subgraph, the numerators contain no more than $n-4$ powers of the loop momentum for that loop \cite{Bern:2012uf}. The exception is that if the one-loop subgraph is a one-loop form factor, one allows the maximal power to be $n-3$. 
	\item  The numerator should preserve the symmetry of the corresponding graph. 
	\item  The numerator should be consistent with the maximal cut \cite{Bern:2007ct} (where all internal legs are cut). One can apply the simple ``rung rule" in such cuts \cite{Bern:1998ug}.
	\end{itemize}

By applying these constraints one obtains an ansatz for the master integrals with a number of free parameters.

\item Given the ansatz for master integrals, one can obtain the numerators of all other integrals in the basis by using a small set of Jacobi relations.  We then impose the same constraints considered above for all integrals. This typically fixes a large number of parameters. It is then necessary to check if the obtained numerators satisfy the full set of color-kinematics equations. If all Jacobi relations are satisfied, we have found a color-dual solution.

\item  It is not guaranteed that the solution obtained is physical and in general free parameters are left at this stage. To both fix these parameters and to make sure that the solution of the integrand is indeed physical, one should check it against a complete set of (preferably D-dimensional) unitarity cuts \cite{Bern:1994zx, Britto:2004nc}. In the next subsection, we discuss how to do these unitarity checks in a systematic way.

\end{enumerate}


\subsubsection{Unitarity checks \label{section-unitarity}}

On a given unitarity cut one can compute the cut integrand both from the ansatz as well as the product of lower-loop amplitudes and form factors, summed over the complete spectrum of the theory under study. For the solution obtained from the procedure outlined above the two expressions must be consistent with each other i.e  
\be \label{unitarity} K_n \sum_{\sigma_n} \sum_i {1\over S_i} \, C_i \, I_i \, \Big|_{\rm cuts}  = \int \prod_i d^4 \eta_{\ell_i} \, {\cal F}^{\rm tree} \prod_a {\cal A}_a^{\rm tree}  \ . \ee
In practice it is convenient to consider color-stripped cuts involving tree amplitudes.  The tree inputs are color-ordered amplitudes or form factors, and the cut integrand on the LHS can be also constructed without considering color factors.

Although the idea is straightforward, in practice there are two technical problems. One is to compute the product of tree amplitudes, where the main challenge is to do the summation over the on-shell states appearing on cut legs. The other is to extract the cut integrand from the ansatz, in particular when there is a large number of basis integrals.


\subsubsection*{Tree input}

We can start from supersymmetric expressions for tree amplitudes and form factors, and take the product of them by integrating out the fermionic variable $\eta_{\ell_i}$ for each cut leg $\ell_i$. For example, in the four-loop computation, we need to consider a quintuple cut which needs the following tree input
\bea && \int \prod_{i=1}^5 d^4 \eta_{l_i}  \Big[ {\cal F}^{\rm MHV}_5(-l_1, - l_2, - l_3, - l_4, - l_5) \, {\cal A}^{\overline{\rm MHV}}_7(p_1, p_2, l_5, l_4, l_3, l_2, l_1) +  \\ && \hskip 2cm  {\cal F}^{\rm NMHV}_5(-l_1, - l_2, - l_3, - l_4, - l_5) \, {\cal A}^{{\rm N}^2{\rm MHV}}_7(p_1, p_2, l_5, l_4, l_3, l_2, l_1) + \nonumber\\ && \hskip 2cm  {\cal F}^{{\rm N}^2{\rm MHV}}_5(-l_1, - l_2, - l_3, - l_4, - l_5) \, {\cal A}^{\rm NMHV}_7(p_1, p_2, l_5, l_4, l_3, l_2, l_1) + \nonumber\\ && \hskip 2cm  {\cal F}^{\textrm{max-non-MHV}}_5(-l_1, - l_2, - l_3, - l_4, - l_5) \, {\cal A}^{\rm MHV}_7(p_1, p_2, l_5, l_4, l_3, l_2, l_1) \Big]  \ . \nonumber \eea
The main complexity is the fermionic integration as it involves high maximally helicity violating (MHV) degrees. This has been discussed for example in \cite{Elvang:2008vz, Bern:2009xq}. In practice we find it convenient to apply the MHV rules as in \cite{Elvang:2008vz}. The supersymmetric form factor was developed in \cite{Brandhuber:2011tv, Bork:2011cj}.

\subsubsection*{Cut integrand from the ansatz}

For the higher-loop integrand given by a large number of integrals, there would be hundreds of cut diagrams contributing to some unitarity cuts.  For example, for the above quintuple cut of the four-loop form factor, there are 192 cut diagrams to sum over. Obviously, one has to develop a systematic method to automatize the computation.  In principle, one could start from the basis graphs, consider various possible cuts of them, and then collect the contributions. In practice, we find the following way more convenient. A similar idea has been used in \cite{Bern:2010tq}.
\begin{enumerate}

\item  Rather than starting from the basis and then doing cuts, we ``inverse" the procedure. We first construct the trivalent tree diagrams of amplitudes and form factors. We take them as the cut tree diagrams on the r.h.s. of (\ref{unitarity}). 

\item Given the trivalent tree graphs, the next step is to construct loop diagrams by sewing trivalent tree diagrams in all possible ways. Since the tree diagrams are trivalent, the loop diagrams obtained are also trivalent. Not all such loop diagrams correspond to basis integrals.  If a loop diagram is isomorphic to one of the basis, then it corresponds to a cut contribution of that basis.   If it is not isomorphic to any basis, we can simply neglect it. In this way, by sewing all possible trivalent diagrams, we are guaranteed to obtain all possible cut contributions from all basis integrals, without double counting. 

\item The cut integrand for each cut diagram is given by the product of the corresponding numerator and tree propagators. In particular, the numerator $N_i$ for a given cut diagram should be re-expressed in terms of external and cut momenta. As we mentioned that it is convenient to consider color-stripped cuts, it is not necessary to consider color factors here%
\footnote{We add a comment on the non-planar cut, which can be obtained by permuting the external legs of tree amplitudes. For these cases it is crucial to obtain the correct sign factor for each cut diagram that is related to the order of tree legs. One way to obtain these sign factors is to assign a color factor for each trivalent vertex of tree graphs, and compare the ``glued" color factor with the color factor of the basis.}. The full cut integrand for a given cut is then given by summing over all cut diagrams
\be K_n \sum_{{\rm cut~graphs}_i} N_i \prod ({\rm tree~propagators}) \ , \ee
which should be compared with the integrand obtained from the product of tree amplitudes.

\end{enumerate}

The above procedure has been applied to a few known results up to three loops and always consistency has been found, which shows that the method is robust. We then apply it to the four-loop form factor, such as the above quintuple cut of four-loop form factors, which also provides consistent results.


\section{Examples  \label{sec:examples}}

In this section the general technology outlined in the previous section is applied to various form factors in ${\cal N}=4$ SYM above one loop%
\footnote{It is easier to work out the one-loop two- and three-point form factors following the higher-loop examples considered here.} which have appeared in the literature. Although the results are known, our techniques illustrate the existence and the power of the duality.


\subsection{Two-point two-loop form factor}

As a warm-up exercise, we consider first the two-loop two-point form factor. This result has been computed by Feynman graph methods in  \cite{vanNeerven:1985ja}.

\begin{figure}[H]
\begin{center}
\includegraphics[height=2.5cm]{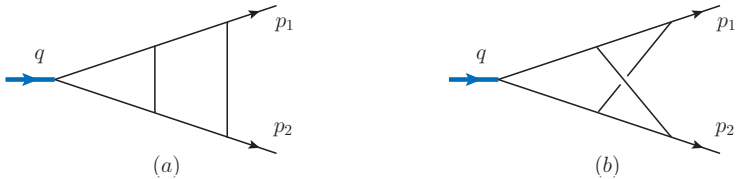}
\caption{\it The integrals for the two-point two-loop form factor.}
\label{fig-F2_2loop}
\end{center}
\end{figure}
First, by equation \eqref{eq:genformfac2pts} the two-point form factor in ${\cal N}=4$ SYM is trivially dependent of the inserted operator through a tree factor, as long as it's in the stress-energy tensor multiplet. This tree factor is factored out, as will be done in every two-point calculation in this article. 

By the rules introduced in the previous section, there are only two trivalent graphs to consider as shown in Figure\,\ref{fig-F2_2loop}: a planar ladder and a non-planar ladder diagram. The Jacobi relations simply tell us that the numerators of both integrals are the same.

By the power counting constraint explained in the previous section, we find that the numerator should be independent of loop momenta. Hence by the kinematics of the problem the numerator should be proportional to a power of $s_{12}$, which will be absorbed into the whole kinematic factor $K_2 = s_{12}^2\, {\cal F}_2^{(0)}$. The numerator is then a purely numerical constant. This numerical constant can be easily fixed by considering any (color-stripped) unitarity cut, which turns out to be one. At the same time, this unitarity cut verifies that nothing has been missed in the construction. In this simple example it is not hard to explicitly compute and verify all possible unitarity cuts, verifying that the result is physical. 

The results including the color and symmetry factors are summarized in Table\,\ref{tab-F2_2loop}. The full form factor result can be obtained as
\bea {\cal F}_2^{(2)} & = & K_2 \,  \sum_{\sigma_2} \sum_{i=a}^b {1\over S_i} \, C_i \, I_i  \\ & = & N_c^2\, \delta^{a_1 a_2}\, s_{12}^2\, {\cal F}^{(0)} \left( 4 \, I_a + I_b \right) \, , \eea
which reproduces exactly the known result \cite{vanNeerven:1985ja}. Note that the color and symmetry factors are responsible for the numerical integer factors, which are $4$ and $1$ for planar and non-planar graphs respectively.

\begin{table}[H]
\begin{center}
\caption{\it The result for the two-point two-loop form factor.
\label{tab-F2_2loop}
}
\vskip .5 cm
\begin{tabular}{l | c | c | c} 
Basis &  Numerator factor  &  Color factor & Symmetry factor \cr \hline
(a) \begin{tabular}{cc}  ~ \\  ~ \end{tabular}   &  1 & $  4 \, N_c^2 \, \delta^{a_1 a_2}  $ & 2 \cr \hline
(b) $\begin{matrix} ~ \\ ~   \end{matrix}$   & 1 & $ 2 \, N_c^2 \, \delta^{a_1 a_2} $ & $4$ \cr \hline
\end{tabular} 
\end{center}
\end{table}


\subsection{Two-point three-loop form factor}

As a more non-trivial example, the two-point form factor at three loops is calculated next by the procedure outlined above. This result has been computed by unitarity methods in \cite{Gehrmann:2011xn}. 

\begin{figure}[H]
\begin{center}
\includegraphics[height=5.5cm]{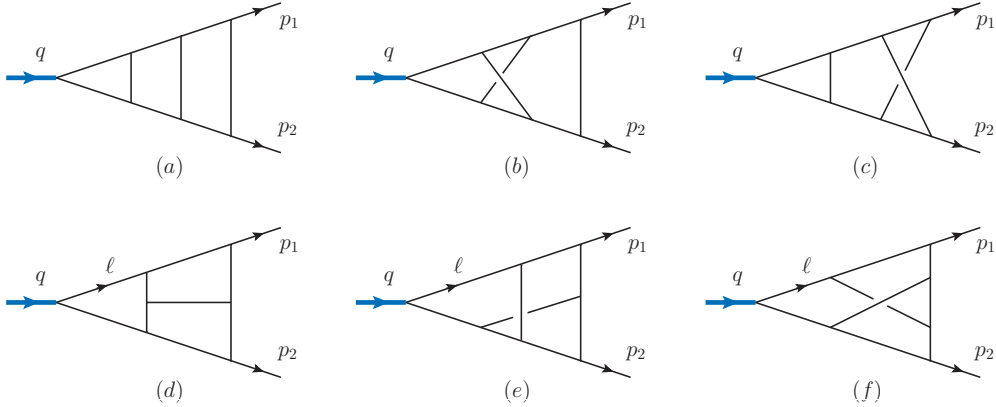}
\caption{\it The integrals for the two-point three-loop form factor.}
\label{fig-F2_3loop}
\end{center}
\end{figure}

First, by generating topologies we can find there are six trivalent diagrams, as shown in Figure \ref{fig-F2_3loop}.%
\footnote{There is one bubble-like graph containing a two-point tree leg which turns out not contribute. For simplicity we do not include it here. In the  four-loop construction, such graphs are as shown in Figure \ref{fig-F2_4loop_e}.}.

By applying the  color-kinematic relation to this set of trivalent diagrams, a set of equations can be obtained for the numerators. It turns out that one can choose the single integral (d) as the master integral. 
One can then make an ansatz for the numerator of this master integral by applying the following three constraints. 
\begin{enumerate}

\item From the power counting property, the numerator should depend only linearly on the loop momentum $\ell$ and there should be no dependence on other loop momenta.  A general ansatz is therefore given as  (note that we have factorized a whole factor $s_{12}^2$)
\be \label{ansatz-d} N_d^{\rm ansatz}(p_1,p_2,\ell) = \alpha_1 \ell \cdot p_1 + \alpha_2 \ell \cdot p_2 + \alpha_3  p_1\cdot p_2  \, , \ee
which contains three parameters $\alpha_i, i=1,2,3$.

\item The numerator should preserve the symmetry of the graph, which implies that it should be invariant under
\be \{ p_1, p_2, \ell \} \quad \Longleftrightarrow \quad \{ p_2, p_1, q - \ell \} \, , \ee
or more explicitly
\be N_d (p_1, p_2, \ell) = N_d(p_2, p_1, q - \ell) \, . \ee
Plugging in the ansatz (\ref{ansatz-d}), we obtain the relation
\be \alpha_2 = - \alpha_1 \, .  \ee

\item Finally, we consider the constraint of maximal cut. From the ``rung rule" we read off the numerator $(\ell - p_1)^2$. On the maximal cut we have
\be  \big[ N_d (p_1, p_2, \ell) - (\ell - p_1)^2 \big] \Big|_{\rm maximal~ cut} = 0  \, .  \ee
This fixes the remaining two parameters 
\be \alpha_1 = -1, \qquad \alpha_3 = -1 \, .  \ee

\end{enumerate}
Therefore, by applying the above constraints we arrive at a unique solution for the master integral
\be N_d^{\rm ansatz} = (p_2-p_1)\cdot \ell -p_1\cdot p_2  \ . \ee

Given this solution, one can check that all Jacobi equations are satisfied.  Other numerators can be obtained from the master integral by using the relations
\bea && N_a = N_b = N_c \, ,  \qquad N_d = - N_e = N_f \, , \\ && N_b(p_1,p_2) = -N_e(p_1,p_2,\ell) - N_e(p_2,p_1,\ell) \ , \eea
where $N_x = N_x(p_1, p_2, \ell)$ if not specified.

\begin{figure}[t]
\begin{center}
\includegraphics[height=4.cm]{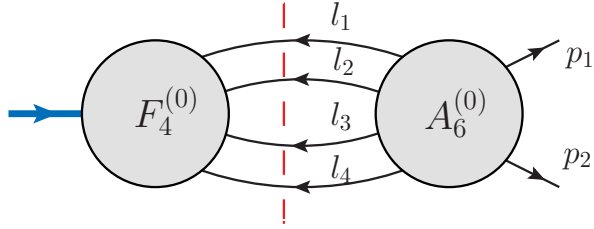}
\caption{\it A quadruple-cut for the two-point three-loop form factor.}
\label{fig-F3_3loop_4cut}
\end{center}
\end{figure}

Now it is essential to check that the solution is indeed physical i.e. satisfies the unitarity cuts. A non-trivial quadruple cut is given as in Figure\,\ref{fig-F3_3loop_4cut}. 
The product of trees is
\bea && \int \prod_{i=1}^4 d^4 \eta_{l_i}  \Big[ {\cal F}^{\rm MHV}_4(-l_1, - l_2, - l_3, - l_4) \, {\cal A}^{\overline{\rm MHV}}_6(p_1, p_2, l_4, l_3, l_2, l_1)+ \\ 
&& \hskip 2cm  {\cal F}^{\rm NMHV}_4(-l_1, - l_2, - l_3, - l_4) \, {\cal A}^{{\rm N}{\rm MHV}}_6(p_1, p_2, l_4, l_3, l_2, l_1) + \nonumber\\ 
&& \hskip 2cm  {\cal F}^{\textrm{max-non-MHV}}_4(-l_1, - l_2, - l_3, - l_4) \, {\cal A}^{\rm MHV}_6(p_1, p_2, l_4, l_3, l_2, l_1) \Big]  \ . \nonumber \eea
From the basis integrals, we obtain the cut integrand as a sum of $29$ cut diagrams. We have compared the two expressions numerically, and have found perfect agreement.

The results have been summarized including the color and symmetry factors in Table\,\ref{tab-F2_3loop}. The full form factor result can be obtained as
\be {\cal F}_2^{(3)} = s_{12}^2 {\cal F}_2^{(0)} \sum_{\sigma_2} \sum_{i=a}^e {1\over S_i} \, C_i \, I_i \ . \ee
Note that the graph (f) has zero color factor and therefore does not contribute to the final result of the form factor. However, it is necessarily involved in solving the Jacobi relations.

\begin{table}[H]
\begin{center}
\caption{\it The result for the two-point three-loop form factor.
\label{tab-F2_3loop}
}
\vskip .5 cm
\begin{tabular}{l | c | c | c} 
Basis &  Numerator factor  &  Color factor & Symmetry factor \cr \hline
(a) \begin{tabular}{cc}  ~ \\  ~ \end{tabular}   &  $ s_{12}^2$ & $  8 \, N_c^3 \, \delta^{a_1 a_2}  $ & 2 \cr \hline
(b) $\begin{matrix} ~ \\ ~   \end{matrix}$   & $s_{12}^2$ & $ 4 \, N_c^3 \, \delta^{a_1 a_2} $ & $4$ \cr \hline
(c) $\begin{matrix} ~ \\ ~   \end{matrix}$   & $s_{12}^2$ & $4  \, N_c^3 \, \delta^{a_1 a_2}$ & $4$ \cr \hline
(d) $\begin{matrix} ~ \\ ~   \end{matrix}$   & $(p_2-p_1)\cdot \ell -p_1\cdot p_2$ & $2  \, N_c^3 \, \delta^{a_1 a_2}$ & 2 \cr \hline
(e) $\begin{matrix} ~ \\ ~   \end{matrix}$   & $- (p_2-p_1)\cdot \ell + p_1\cdot p_2$ & $2 \, N_c^3 \, \delta^{a_1 a_2}$ & 1 \cr \hline
(f) $\begin{matrix} ~ \\ ~   \end{matrix}$    & $(p_2-p_1)\cdot \ell -p_1\cdot p_2$ & 0 & 2 \\ \hline
\end{tabular} 
\end{center}
\end{table}

The result  we obtain by applying color-kinematic duality  seems quite different from that in \cite{Gehrmann:2011xn}.  However, it is a simple check that the results are equivalent, by using the identities given in section 3 of \cite{Gehrmann:2011xn} between different integrals.   Our result is presented in a much simpler form which involves only trivalent graphs. The numerical integer factors which have no easy interpretation in \cite{Gehrmann:2011xn} are also naturally explained here by the color and symmetry factors.


\subsection{Three-point two-loop form factor}

As an example containing more external legs, we present the three-point two-loop form factor in a form satisfying all Jacobi equations. The three-point two-loop form factor was computed in \cite{Brandhuber:2012vm}. 

\begin{figure}[h]
\begin{center}
\includegraphics[height=5.5cm]{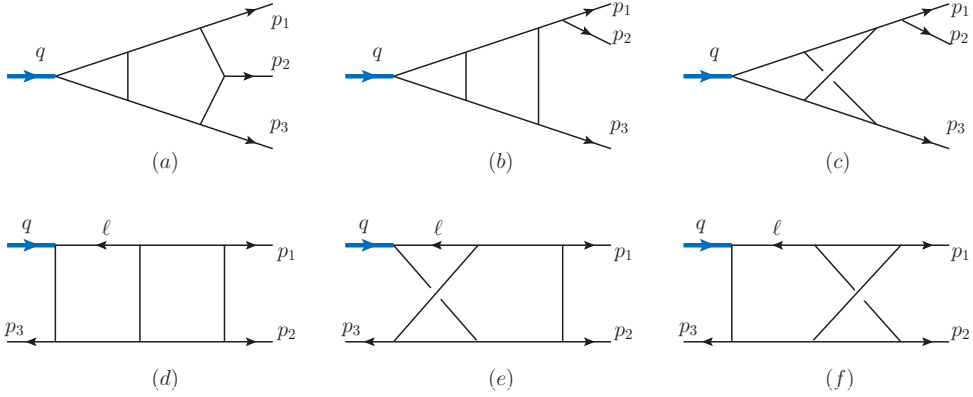}
\caption{\it The integrals for the three-point two-loop form factor.}
\label{fig-F3_2loop}
\end{center}
\end{figure}

There are six  trivalent diagrams, as  shown in Figure. \ref{fig-F3_2loop}. 

By inspecting the set of Jacobi relations, we can find that one needs at least two master integrals, which may be chosen as  graph (a) and graph (d). The numerators of the other graphs can be obtained by the following relations
\bea && N_b(p_1,p_2,p_3) = N_a(p_2,p_3,p_1) + N_a(p_3,p_2,p_1) , \\
&& N_e(p_1,p_2,p_3, \ell) = N_a(p_1,p_2,p_3) - N_d(p_1,p_2,p_3, \ell) , \\ 
&&  N_c = N_b , \qquad\qquad N_f = N_d  \, ,
 \eea
where $N_x = N_x (p_1, p_2, p_3, \ell)$ if not specified.

\begin{table}[h]
\begin{center}
\caption{\it The result for the three-point two-loop form factor.
\label{tab-F3_2loop}
}
\vskip .5 cm
\begin{tabular}{l | c | c | c} 
Basis &  Numerator factor  &  Color factor & Symmetry factor \cr \hline
(a) \begin{tabular}{cc}  ~ \\  ~ \end{tabular}   &  $ q^2 s_{12} s_{23}/2$ & $  2 \, N_c^2 \, \tilde f^{a_1a_2a_3}  $ & 2 \cr \hline
(b) $\begin{matrix} ~ \\ ~   \end{matrix}$   & $q^2 s_{12} (s_{13}+s_{23})/2$ & $ 2  \, N_c^2 \, \tilde f^{a_1a_2a_3} $ & $2$ \cr \hline
(c) $\begin{matrix} ~ \\ ~   \end{matrix}$   & $q^2 s_{12} (s_{13}+s_{23})/2$ & $2  \, N_c^2 \, \tilde f^{a_1a_2a_3}$ & 4 \cr \hline
(d) $\begin{matrix} ~ \\ ~   \end{matrix}$   & $s_{12} (s_{13} \ell \cdot p_1 - s_{23} \ell \cdot p_2)$ & $  \, N_c^2 \, \tilde f^{a_1a_2a_3}$ & 1 \cr \hline
(e) $\begin{matrix} ~ \\ ~   \end{matrix}$   & $q^2 s_{12} s_{23}/2 - s_{12} (s_{13} \ell \cdot p_1 - s_{23} \ell \cdot p_2)$ & $  \, N_c^2 \, \tilde f^{a_1a_2a_3}$ & 2 \cr \hline
(f) $\begin{matrix} ~ \\ ~   \end{matrix}$    & $s_{12} (s_{13} \ell \cdot p_1 - s_{23} \ell \cdot p_2)$ & 0 & 2 \\ \hline
\end{tabular} 
\end{center}
\end{table}

Using the numerators in \cite{Brandhuber:2012vm} for two master integrals, one can check that all the Jacobi relations are automatically satisfied.  The numerator solution is given in Table \ref{tab-F3_2loop}. The color and symmetry factors are also given in Table\,\ref{tab-F3_2loop}. The full result can be constructed from the table as 
\be {\cal F}_3^{(2)} = {\cal F}_3^{(0)} \sum_{\sigma_3} \sum_{i=a}^e {1\over S_i} \, C_i \, I_i  \ . \ee
It is  a simple check that the result is equivalent to the one presented in \cite{Brandhuber:2012vm}.  We have also checked that the solution passes various triple cuts following the procedure of section\,\ref{section-unitarity}.


\section{Four-loop two-point form factor \label{sec:fourloop}}

In this section the full four-loop integrand of the two-point form factor is constructed. To our knowledge this has never been computed before. 
\begin{figure}[h]
\begin{center}
\subfigure{
\includegraphics[height=8.5cm]{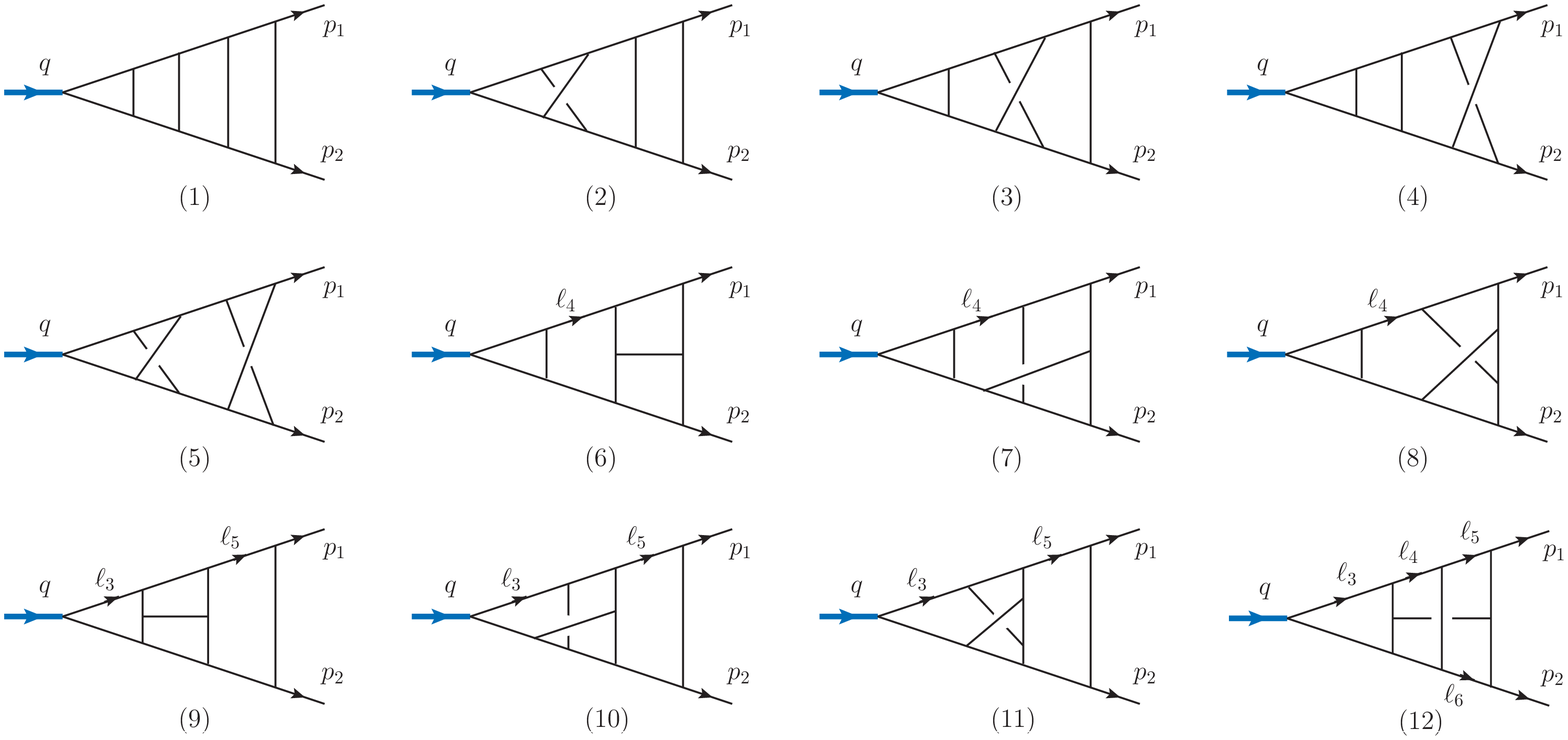}
}
\vskip .2cm
\subfigure{~~\,
\includegraphics[height=4.62cm]{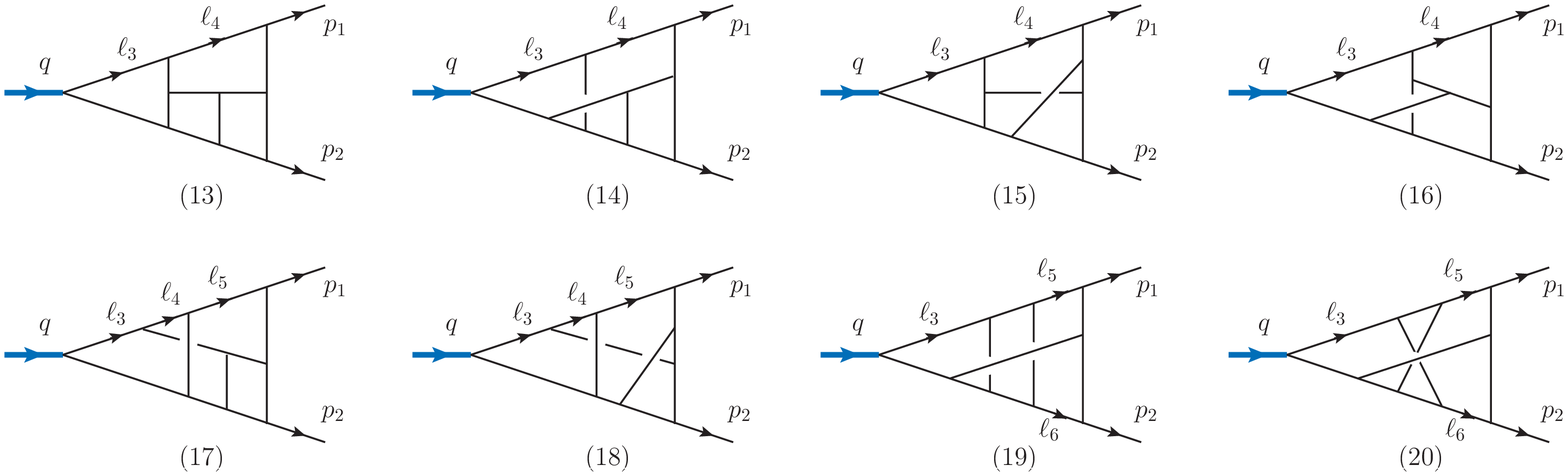}
}
\caption{\it The basis (1)$-$(20) of two-point four-loop form factor. These diagrams have only planar contributions. Six of them have zero color factor.}
\label{fig-F2_4loop_a}
\end{center}
\end{figure}

\begin{figure}[h]
\begin{center}
\subfigure{~~
\includegraphics[height=7.1cm]{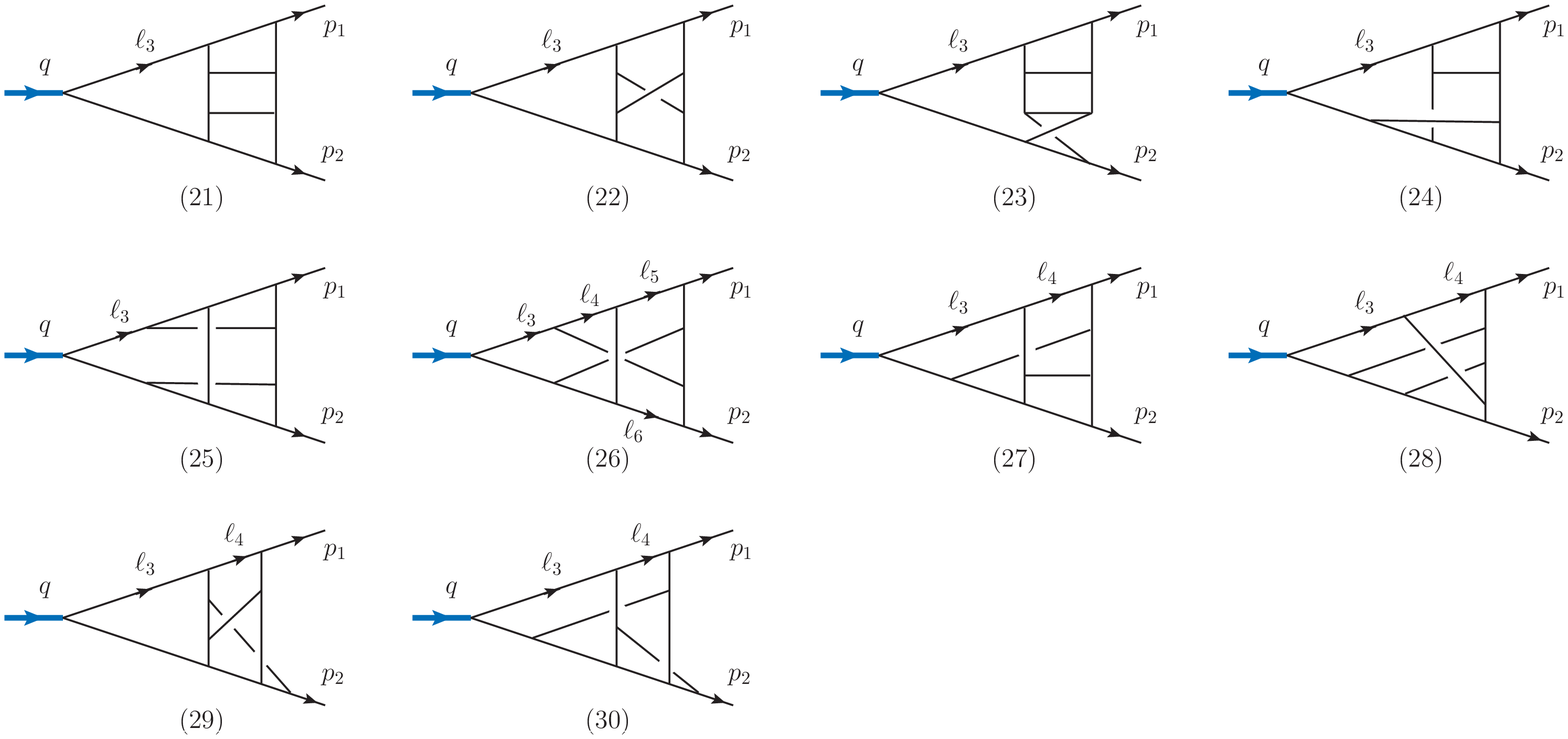}
}
\vskip .3cm
\subfigure{~~
\includegraphics[height=2.0cm]{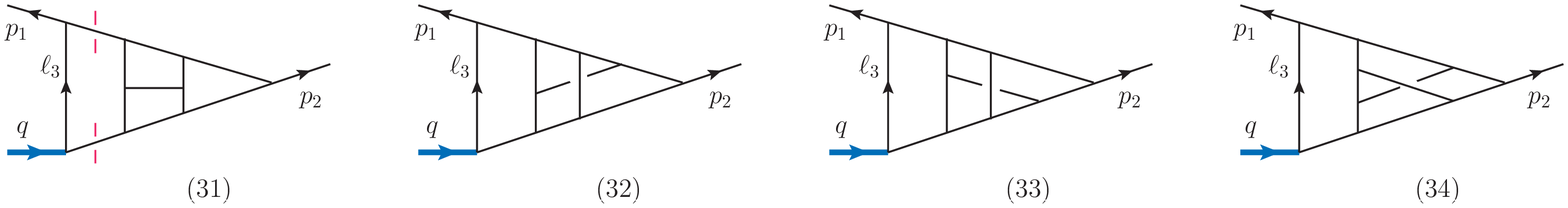}
}
\caption{\it The basis (21)$-$(34) of two-point four-loop form factor. These 14 integrals have non-planar contributions. Five of them also have planar contributions.}
\label{fig-F2_4loop_d}
\end{center}
\end{figure}

\begin{figure}[h]
\begin{center}
\includegraphics[height=2.2cm]{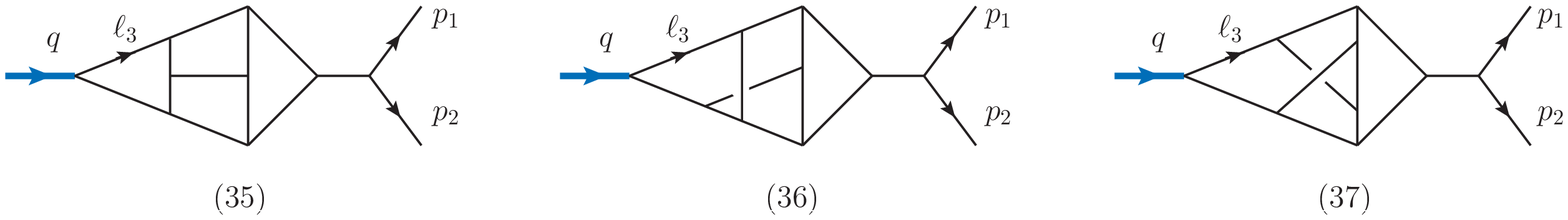}
\caption{\it The three bubble integrals which do not contribute to the final result.}
\label{fig-F2_4loop_e}
\end{center}
\end{figure}


\subsection{Outline of the construction}

The construction is similar to the three-loop example. Here we outline the main steps.

\begin{enumerate}

\item  
There are in total $37$ trivalent diagrams to consider, which are  as given in Figures \ref{fig-F2_4loop_a}$-$\ref{fig-F2_4loop_e}.  For completeness, we consider also three bubble-like graphs as shown in Figure\,\ref{fig-F2_4loop_e}.

\item
By applying the Jacobi relations for all possible propagators of all 37 integrals, one obtains a set of equations for the numerator functions. Part of the equations are given in appendix\,\ref{jacobi-4loop}.
From the equations, one can identify two integrals as the master integrals:  (13) and (21), which both have planar topology.

\item
We then construct an ansatz for the numerators of the master integrals. By power-counting, one can write down a most general ansatz as a polynomial of loop and external momenta for the numerator. For the master integral (21), the numerator depends at most quadraticly on $\ell_3$, so a general ansatz is 
\bea N_{21} &=&  \alpha_1 (\ell_3 \cdot p_1)^2 + \alpha_2 (\ell_3 \cdot p_2)^2 + \alpha_3 (\ell_3\cdot p_1) (\ell_3 \cdot p_2) \nonumber\\ && + \, \big[ \alpha_4 (\ell_3 \cdot \ell_3) + \alpha_5 (\ell_3 \cdot p_1) + \alpha_6 (\ell_3 \cdot p_2) + \alpha_7 (p_1 \cdot p_2) \big] (p_1 \cdot p_2) \, .  \eea
For the master integral (13), the numerator depends linearly on $\ell_3$ and $\ell_4$, and a general ansatz is
\bea N_{13} &=&  \beta_1 (\ell_3 \cdot p_1)(\ell_4\cdot p_1) + \beta_2 (\ell_3 \cdot p_2)(\ell_4\cdot p_2) + \beta_3 (\ell_3 \cdot p_1)(\ell_4\cdot p_2) + \beta_4 (\ell_3 \cdot p_2)(\ell_4\cdot p_1) \nonumber\\ && \hskip -1.4cm + \, \big[ \beta_5 (\ell_3 \cdot \ell_4) + \beta_6 (\ell_3 \cdot p_1) + \beta_7 (\ell_3 \cdot p_2) + \beta_8 (\ell_4 \cdot p_1) + \beta_9 (\ell_4 \cdot p_2) + \beta_{10} (p_1 \cdot p_2) \big] (p_1 \cdot p_2) \nonumber .\eea
There are in total $17$ parameters.  The numerators of all other basis elements can be determined from these two via Jacobi relations.

\item Then we can impose the constraints from the maximal cut%
\footnote{We apply the maximal cut and symmetry constraints for master integrals and other integrals together.}. 
There are five planar integrals for which one can easily read off the numerators directly from the rung rule:
\bea N_1^{\rm rr}~ & = &  s_{12}^2 \, , \\  
N_6^{\rm rr}~ &=&  s_{12}\, (\ell_4 - p_1)^2 \, , \\
N_9^{\rm rr}~ &=&  s_{12}\, (\ell_3- \ell_5)^2 \, , \\ 
N_{13}^{\rm rr}~ &=&  (\ell_3 - p_1)^2 (\ell_4 - p_1 - p_2)^2 \, , \\ 
N_{21}^{\rm rr}~ &=&  (\ell_3 - p_1)^4 \, .  \eea
In the maximal cut the ansatz must be consistent with these coefficients i.e.
\be \big(N_i - N_i^{\rm rr} \big) \big|_{\rm maximal~cut}  = 0 \, . \ee
These equations fix ten parameters.

\item
The numerators should satisfy the symmetry of the diagrams.
Since the numerators of many graphs are related to each other by Jacobi relations, it is enough to consider the symmetries of the following graphs
\be \{\, (1), \, (6) , \, (9) , \, (12) , \, (20) , \, (21) , \, (25) , \, (26)  \, \} \ .
\ee
For example graph (26) has the symmetry:
\be \{p_1, p_2, \ell_3, \ell_4, \ell_5, \ell_6 \}  \ \leftrightarrow \ 
\{ p_2, p_1, p_{12} - \ell_3, \ell_5 + \ell_6 - \ell_4, \ell_6, \ell_5 \} \, . \ee
These symmetry constraints can fix $4$ parameters. Together with the above constraints from the maximal cut,  one can fix in total 13 parameters, which can be explicitly given as
\bea && \alpha_2 = \alpha_1 , \qquad \alpha_3  = -4 + 2 \alpha_1 , \qquad \alpha_5 = 4- 2 \alpha_1 - \alpha_4 , \qquad \alpha_6 = - 2 \alpha_1 - \alpha_4 , \nonumber\\ && \alpha_7 = \alpha_1 + \alpha_4 , \qquad \beta_3 = 4 + \beta_2 , \qquad \beta_4 = -4 + \beta_1 , \nonumber\\ && \beta_5 = \beta_6= -\beta_7 = \beta_{10}= -2, \qquad \beta_8 = 2 - \beta_1, \qquad \beta_9 = - \beta_2 \, .  \eea
Substituting these solutions into the numerators, we find that the full set of Jacobi equations are satisfied. As a consistency check, all the numerators  have the required power counting behavior. It also turns out that the numerators of the bubble-like integrals (35)$-$(37) are zero.

\item

There is another simple cut one can apply for the basis integrals (31)$-$(34), as shown in graph  (31) in  Figure \ref{fig-F2_4loop_d}.  The cut integrand involves a three-point loop amplitude, which implies that the numerator should vanish under this cut. This simple constraint fixes two more parameters:
\be \beta_1 = - \alpha_1 , \qquad \beta_2 = 2 - {\alpha_4 \over 2} \, . \ee

\end{enumerate}
Therefore, we arrive at a solution of numerators which depends on only two parameters \{$\alpha_1, \alpha_4$\}.


\subsection{Unitarity checks}

\begin{figure}[h]
\begin{center}
\includegraphics[height=4.cm]{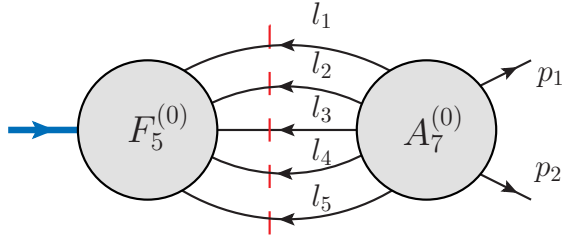}
\caption{A quintuple cut for the two-point four-loop form factor.}
\label{fig-F2_4loop_5cut}
\end{center}
\end{figure}

To make sure that the solution is indeed the form factor result, we perform a large number of unitarity cuts. In particular a non-trivial quadruple cut is given as in Figure\,\ref{fig-F2_4loop_5cut}. From the basis integrals, we obtain the cut integrand in terms of 192 cut diagrams from the following 18 integrals:
\be \{\, (1), \, (2) , \, (3) , \, (4) , \, (5) , \,(6), \, (7) , \, (9) , \, (10) , \, (12) , \, (13) , \, (14), \, (17) , \,(19),  \, (21) , \,(25),  \, (30) , \,  (31) \, \} \, .
\ee
This should be compared with the product of trees, which involves non-trivial next-to-next-to-MHV tree amplitudes
\bea && \int \prod_{i=1}^5 d^4 \eta_{l_i}  \Big[ {\cal F}^{\rm MHV}_5(-l_1, - l_2, - l_3, - l_4, - l_5) \, {\cal A}^{\overline{\rm MHV}}_7(p_1, p_2, l_5, l_4, l_3, l_2, l_1)+  \\ && \hskip 2cm  {\cal F}^{\rm NMHV}_5(-l_1, - l_2, - l_3, - l_4, - l_5) \, {\cal A}^{{\rm N}^2{\rm MHV}}_7(p_1, p_2, l_5, l_4, l_3, l_2, l_1)+ \nonumber\\ && \hskip 2cm  {\cal F}^{{\rm N}^2{\rm MHV}}_5(-l_1, - l_2, - l_3, - l_4, - l_5) \, {\cal A}^{\rm NMHV}_7(p_1, p_2, l_5, l_4, l_3, l_2, l_1) + \nonumber\\ && \hskip 2cm  {\cal F}^{\textrm{max-non-MHV}}_5(-l_1, - l_2, - l_3, - l_4, - l_5) \, {\cal A}^{\rm MHV}_7(p_1, p_2, l_5, l_4, l_3, l_2, l_1) \Big]  \ . \nonumber \eea
The two results are compared numerically to very high precision. We find that the remaining two parameters should satisfy the relation
\be 6 = \alpha_1 + {7\over2} \alpha_4 \, . \label{2para-relation} \ee

We should emphasize that this is a very non-trivial consistency check for our result, considering that both the cut integrand and the product of trees are very different and complicated expressions. The two sides are however identical to each other if and only if the very simple relation (\ref{2para-relation}) holds. This leaves one free parameter in the solution which is written below in full detail. The numerators of the two master integrals can be given explicitly as
\bea  N_{13} &=&  - (\ell_3 \cdot p_1)(\ell_4\cdot p_1) - (\ell_3 \cdot p_2)(\ell_4\cdot p_2) -5 (\ell_3 \cdot p_1)(\ell_4\cdot p_2) + 3 (\ell_3 \cdot p_2)(\ell_4\cdot p_1) \nonumber\\ && + \, (p_1 \cdot p_2)\big[ 2 (\ell_3 \cdot \ell_4) + 2 (\ell_3 \cdot p_1) - 2 (\ell_3 \cdot p_2) - (\ell_4 \cdot p_1) +  (\ell_4 \cdot p_2) + 2 (p_1 \cdot p_2) \big]  \\ && - {1\over7}(\alpha_1 + 1) (\ell_3 \cdot p_{12} - p_1 \cdot p_2)(\ell_4 \cdot (7p_1 - p_2)) \, , \label{N13} \nonumber \\
 N_{21}  &=&  - (\ell_3 \cdot p_1)^2 - (\ell_3 \cdot p_2)^2 -6 (\ell_3\cdot p_1) (\ell_3 \cdot p_2) + (p_1 \cdot p_2)\big[ 2 (\ell_3 \cdot \ell_3) + 4 (\ell_3 \cdot p_1) + (p_1 \cdot p_2) \big]  \nonumber\\ && + (\alpha_1 + 1) \big[ (\ell_3 \cdot p_{12} - p_1 \cdot p_2)^2 - {2\over7} (\ell_3\cdot (\ell_3 - p_{12}) + p_1\cdot p_2)(p_1\cdot p_2) \big] \,  ,  \label{N21} \eea
where we separate the dependence on the remaining parameter into a term proportional to $(\alpha_1 +1)$ which would be zero for the possible choice of (\ref{choice2para}). Using the Jacobi relations in Appendix\,\ref{jacobi-4loop}, one can obtain all other numerators. They are explicitly given in Table \ref{tab-F2_4loop}, and are also written in the form of parameter independent terms plus a term proportional to $(\alpha_1+1)$. Since up to now all coefficients in the numerators have been integers, it is plausible that 
\be \alpha_1 +1 = 7 k  \qquad k \in \mathbb{Z} \ee
which achieves this. Moreover, the most natural choice is $\alpha_1 = -1$ which would simplify the result considerably. This leads to the natural values of the two parameters 
\be \alpha_1 = -1, \qquad \alpha_4 = 2 \, . \label{choice2para} \ee
Actually, the possibility to find only integer coefficients in the ansatz is a result in itself. 

\begin{figure}[t]
\begin{center}
\includegraphics[height=6.8cm]{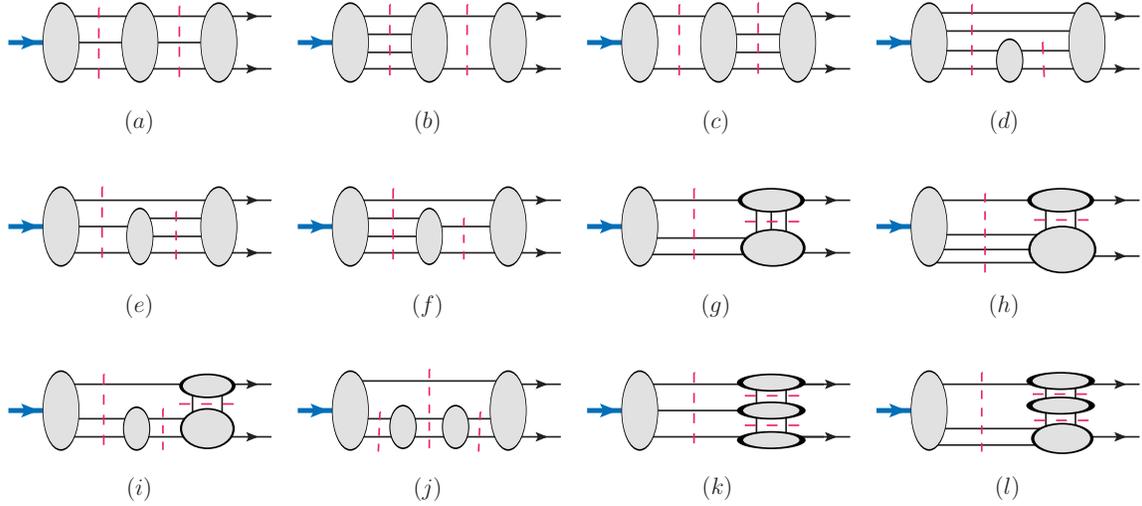}
\caption{\it Other cuts that have been checked. A few non-planar cuts which are obtained by permuting the legs of tree sub-amplitudes have also been checked.}
\label{fig-F2_4loop_multicuts}
\end{center}
\end{figure}

Some other cuts have also been checked as shown in Figure\,\ref{fig-F2_4loop_multicuts}, which are all consistent with the above solution.  A few of them require the same constraint (\ref{2para-relation}), however, none of them provides any new constraint for the parameters. In particular, ordinary unitarity cuts which isolate a tree level two-point form factor cannot fix the coefficient. Therefore we finally arrive at a one-parameter solution of the integrand, as far as the many non-trivial constraints we have imposed.

\subsection{Full result}

The explicit results including  the color%
\footnote{To be explicit, we write the color factor for the $SU(N_c)$ gauge group. To translate to general groups one can rewrite our results in terms of the quadratic Casimir $C_A$ and the quartic Casimir $d_{44}$ as outlined in appendix \ref{app:color}. } 
and symmetry factors are summarized  in Table \ref{tab-F2_4loop}. The full four-loop form factor can be obtained as
\be {\cal F}_2^{(4)} = s_{12}^2 {\cal F}_2^{(0)} \sum_{\sigma_2} \sum_{i=1}^{34} {1\over S_i} \, C_i \, I_i \ . \ee
Note that six of them have zero color factor and do not contribute to the final answer. However, they do play an important role in the above construction%
\footnote{As a side comment, we mention that these bases with zero color factor actually have non-trivial kinematic contributions in some of the non-planar unitarity cuts we have checked. Therefore they in some sense do contain some physical information even though they do not contribute to the final result. If an observable in gravity would exist, we would expect these terms to play a non-trivial role there. }.

A few comments on the result are in order.
\begin{itemize}

\item

All the $34$ integrals have the maximal number of loop propagators. There are no multi-point tree legs such as the snail-like diagrams appearing in the case of four-loop four-point amplitudes \cite{Bern:2012uf}. This seems to be a special feature for the two-point form factor, and may be also true at higher loops.

\item 
Our construction is based on color-kinematic duality which does not rely on the dimension of loop momentum variables. On the other hand, in the unitarity construction one usually needs to worry about cut non-constructible part such as the  $\mu^2$ terms which may be missed in four-dimensional unitarity. It is reasonable to believe that our result should already contain such information and is true for general dimension.  A full set of $D$-dimensional unitarity checks would be interesting to perform.

\item
All the numerators depend at most quadraticly on loop momenta. By naive power-counting this implies that the four-loop result has no UV divergence below the critical dimension of $D = 5{1\over2}$, as expected. Note that there could be further cancellations beyond this hidden in the integrals. 

\item 
The integrand solution we have obtained still depends on one parameter. In all the unitarity cuts we have performed, this parameter is not fixed although as argued above it does have a natural value%
\footnote{Although not likely, it might happen that this is a real freedom at the integrand level, which means that after integration the dependence will vanish.}. It may be uniquely fixed by some other more sophisticated cuts yet to be performed, which should also provide further consistency checks for our result. Another important check will follow from the evaluation of the integrals and their infrared singularity structure \cite{inprogress}.

\end{itemize}

\begin{table}[H]
\begin{center}
\caption{The result for the two-point four-loop form factor.  
\label{tab-F2_4loop}
}
\vskip .5 cm
\begin{tabular}{l | c | c | c} 
Graph &  Numerator factor  &  Color factor & \begin{tabular}{c}  Symmetry \\  factor \end{tabular} \cr \hline
\hline
(1) \begin{tabular}{cc}  ~ \\  ~ \end{tabular}   &  $ s_{12}^2$ & $  16 \, N_c^4 \, \delta_{a_1 a_2}  $ & 2 \cr \hline
(2) $\begin{matrix} ~ \\ ~   \end{matrix}$   & $N_1 $ & $ 8\, N_c^4  \, \delta_{a_1 a_2} $ & $4$ \cr \hline
(3) $\begin{matrix} ~ \\ ~   \end{matrix}$   & $N_1 $ & $ 8\, N_c^4  \, \delta_{a_1 a_2} $ & $4$ \cr \hline
(4) $\begin{matrix} ~ \\ ~   \end{matrix}$   & $N_1 $ & $ 8\, N_c^4  \, \delta_{a_1 a_2} $ & $4$ \cr \hline
(5) $\begin{matrix} ~ \\ ~   \end{matrix}$   & $N_1 $ & $ 4\, N_c^4  \, \delta_{a_1 a_2} $ & $8$ \cr \hline
\hline
(6) $\begin{matrix} ~ \\ ~   \end{matrix}$   & $- s_{12} [ \ell_4\cdot (p_1 - p_2) + p_1 \cdot p_2 ] $ & $ 4\, N_c^4  \, \delta_{a_1 a_2} $ & $2$ \cr \hline
(7) $\begin{matrix} ~ \\ ~   \end{matrix}$   &  $- N_6$ & $ 4\, N_c^4 \, \delta_{a_1 a_2} $ & $1$ \cr \hline
(8) $\begin{matrix} ~ \\ ~   \end{matrix}$   & $N_6$ & 0 & $2$ \cr \hline
\hline
(9) $\begin{matrix} ~ \\ ~ \\ ~   \end{matrix}$   & 
 \begin{tabular}{l}  $2 (\ell_3\cdot p_{12} ) (\ell_5\cdot p_{12} ) - s_{12} (2 \ell_3\cdot\ell_5 + p_1 \cdot p_2) $ \\ $ - {6\over7}(\alpha_1 + 1) (\ell_3 \cdot p_{12} - p_1 \cdot p_2)  (\ell_5 \cdot p_{12} - p_1 \cdot p_2) $
 \end{tabular} 
 & $ 4\, N_c^4  \, \delta_{a_1 a_2}$ & $2$ \cr \hline
(10) $\begin{matrix} ~ \\ ~   \end{matrix}$   & $-N_9$ & $ 4\, N_c^4  \, \delta_{a_1 a_2}$ & $1$ \cr \hline
(11) $\begin{matrix} ~ \\ ~   \end{matrix}$   & $N_9$  & $0$ & $2$ \cr \hline
\hline
(12) \begin{tabular}{cc}  ~ \\  ~ \\ ~\\~\\~ \\ ~ \\ ~  \end{tabular}   &   
 \begin{tabular}{l} $ - (\ell_3 \cdot p_1)\big[ \ell_5\cdot (p_1+5 p_2) - \ell_6 \cdot (p_1 - 3 p_2) \big]  $\\  $ +  (\ell_3 \cdot p_2) \big[ \ell_5\cdot (3 p_1 - p_2) + \ell_6 \cdot (5 p_1+ p_2) \big] $\\  $ + (p_1 \cdot p_2)\big[ 2 \ell_3 \cdot (p_1 - p_2 + \ell_5 - \ell_6)$ \\  $ - 3 (\ell_5 + \ell_6) \cdot (p_1- p_2) + 2 \ell_4 \cdot (p_1 - p_2) + s_{12} \big]  $ \\ $ + {1\over7}(\alpha_1 + 1) (\ell_3 \cdot p_{12} - p_1 \cdot p_2) $ \\   \quad $ \times [\ell_5 \cdot (7p_1 - p_2) + \ell_6 \cdot (p_1 - 7 p_2)] $ \end{tabular} 
  & $  2\, N_c^4 \, \delta_{a_1 a_2}  $ & 2 \cr \hline
\hline
(13) \begin{tabular}{cc}  ~ \\ ~\\ ~\\ ~\\  ~ \\ ~ \end{tabular}   & 
\begin{tabular}{l} $  (\ell_3 \cdot p_1)(\ell_4\cdot (p_1+5 p_2))  $\\ $ - (\ell_3 \cdot p_2) ( (\ell_4\cdot (3 p_1 - p_2)) $ \\   $ - (p_1 \cdot p_2)\big[  2 \ell_3 \cdot (\ell_4+ p_1-p_2)$ \\  $ - \ell_4 \cdot (p_1- p_2) + s_{12} \big]  $ \\ $  - {1\over7}(\alpha_1 + 1) (\ell_3 \cdot p_{12} - p_1 \cdot p_2)(\ell_4 \cdot (7p_1 - p_2)) $  \end{tabular} 
 & $  2 \, N_c^4 \, \delta_{a_1 a_2}  $ & 1 \cr \hline
\end{tabular} 
\end{center}
\end{table}

\newpage 
\begin{center}
\begin{tabular}{ c } 
{\bf Table 5  (continued)}. The result for the two-point four-loop form factor.
\end{tabular}
\end{center}

\begin{center}
%
\begin{tabular}{l | c | c | c} 
Graph &  Numerator factor  &  Color factor & \begin{tabular}{c}  Symmetry \\  factor \end{tabular} \cr \hline
\hline
(14) \begin{tabular}{cc}  ~ \\  ~ \end{tabular}   &  $-N_{13}$ & $  2\, N_c^4 \, \delta_{a_1 a_2}  $ & 1 \cr \hline
(15) \begin{tabular}{cc}  ~ \\  ~ \end{tabular}   &  $-N_{13}$ & $ 0$ & 1 \cr \hline
(16) \begin{tabular}{cc}  ~ \\  ~ \end{tabular}   &  $N_{13}$ & $ 0  $ & 1 \cr \hline 
\hline
(17) \begin{tabular}{cc}   ~ \\ ~\\ ~\\ ~\\  ~ \\ ~ \end{tabular}   & 
  \begin{tabular}{l} $ - (\ell_3 \cdot p_1)(\ell_5\cdot (p_1+5 p_2))  $\\  $ +  (\ell_3 \cdot p_2)(\ell_5\cdot (3 p_1 - p_2)) $\\  $ + (p_1 \cdot p_2)\big[ 2 (\ell_3 \cdot \ell_5) + 2 \ell_4 \cdot (p_1-p_2)$ \\  $ - 3 \ell_5 \cdot (p_1- p_2)  \big]   + {1\over7}(\alpha_1 + 1)\times  $ \\  \quad $  (\ell_3 \cdot p_{12} - p_1 \cdot p_2)(\ell_5 \cdot (7p_1 - p_2)) $ \end{tabular} 
   & $  2\, N_c^4 \, \delta_{a_1 a_2}  $ & 1 \cr \hline
(18) \begin{tabular}{cc}  ~ \\  ~ \end{tabular}   &  $-N_{17}$ & $ 0$ & 2 \cr \hline
\hline
(19) \begin{tabular}{cc}   ~ \\ ~\\ ~\\ ~\\  ~ \\ ~ \\ ~ \end{tabular}   &  
 \begin{tabular}{l} $ (\ell_3 \cdot p_1)\big[ \ell_5\cdot (p_1+5 p_2) - \ell_6 \cdot (p_1 - 3 p_2) \big]  $\\  $ -  (\ell_3 \cdot p_2) \big[ \ell_5\cdot (3 p_1 - p_2) + \ell_6 \cdot (5 p_1+ p_2) \big] $\\  $ - (p_1 \cdot p_2)\big[ 2 \ell_3 \cdot (p_1 - p_2 + \ell_5 - \ell_6)$ \\  $ - 3 (\ell_5 + \ell_6) \cdot (p_1- p_2)  \big]  $ \\ $ - {1\over7}(\alpha_1 + 1) (\ell_3 \cdot p_{12} - p_1 \cdot p_2) $ \\   \quad $ \times [\ell_5 \cdot (7p_1 - p_2) + \ell_6 \cdot (p_1 - 7 p_2)] $ \end{tabular}  
  & $  2\, N_c^4 \, \delta_{a_1 a_2}  $ & 1 \cr \hline
(20) \begin{tabular}{cc}  ~ \\  ~ \end{tabular}   &  $N_{19}$ & $ 0$ & 2 \cr \hline
\hline
(21)  $\begin{matrix} ~ \\ ~ \\ ~ \\ ~ \\~  \end{matrix}$ & 
\begin{tabular}{c} $- (\ell_3 \cdot p_1)^2 - (\ell_3 \cdot p_2)^2 -6 (\ell_3\cdot p_1) (\ell_3 \cdot p_2)$ \\ $+  (p_1 \cdot p_2)\big[ 2 (\ell_3 \cdot \ell_3) + 4 (\ell_3 \cdot p_1) + p_1 \cdot p_2 \big]  $ \\ $ + (\alpha_1 + 1) \big[ (\ell_3 \cdot p_{12} - p_1 \cdot p_2)^2 $ \\ $  - {2\over7} (\ell_3\cdot (\ell_3 - p_{12}) + p_1\cdot p_2)(p_1\cdot p_2) \big]  $\end{tabular} 
 & $ (2 \,N_c^4 + 24\, N_c^2) \, \delta_{a_1 a_2} $ & $2$ \cr \hline
\end{tabular} 
\end{center}


\newpage 
\begin{center}
\begin{tabular}{c } 
{\bf Table 5 (continued)}. The result for the two-point four-loop form factor.
\end{tabular}
\end{center}

\begin{center}
%
\begin{tabular}{l | c | c | c} 
Graph &  Numerator factor  &  Color factor & \begin{tabular}{c}  Symmetry \\  factor \end{tabular} \cr \hline
\hline
(22) $\begin{matrix} ~ \\ ~   \end{matrix}$   & $N_{21}$ & $  24 \,N_c^2 \, \delta_{a_1 a_2} $ & $2$ \cr \hline
(23) $\begin{matrix} ~ \\ ~   \end{matrix}$   & $N_{21}$ & $  24 \, N_c^2 \, \delta_{a_1 a_2} $ & $2$ \cr \hline
(24) $\begin{matrix} ~ \\ ~   \end{matrix}$   & $N_{21}$ & $  24 \, N_c^2 \, \delta_{a_1 a_2} $ & $1$ \cr \hline
(25) $\begin{matrix} ~ \\ ~   \end{matrix}$   & $N_{21}$ & $ (2\, N_c^4 + 24 \, N_c^2) \, \delta_{a_1 a_2} $ & $4$ \cr \hline
\hline
(26) $\begin{matrix}  ~ \\ ~\\ ~\\ ~\\  ~ \\ ~ \\ ~ \\~\\~\\~ \\~\end{matrix}$   & 
\begin{tabular}{c} $  -(\ell_3\cdot p_1)^2 -(\ell_3\cdot p_2)^2 - 6(\ell_3\cdot p_1) (\ell_3\cdot p_2) $ \\ $ + (\ell_3\cdot p_1) \big[ \ell_5\cdot (p_1 + 5 p_2) - \ell_6 \cdot (p_1-3p_2) \big]  $ \\ $ - (\ell_3 \cdot p_2) \big[ \ell_5 \cdot ( 3p_1 - p_2) + \ell_6 \cdot (5 p_1 + p_2) \big]  $ \\ $ + (p_1 \cdot p_2) \big[ 2 \ell_3 \cdot (p_{12} + \ell_3 - \ell_5 + \ell_6)  $\\$  + 3 (\ell_5 + \ell_6) \cdot (p_1 - p_2 ) $  \\ $ -2  \ell_4 \cdot (p_1 - p_2) - p_1 \cdot p_2  \big]   $ \\ $ + (\alpha_1 + 1) \big\{ (\ell_3 \cdot p_{12} - p_1 \cdot p_2)^2 $ \\ $  - {2\over7} (\ell_3\cdot (\ell_3 - p_{12}) + p_1\cdot p_2)(p_1\cdot p_2)  $  \\ $ - {1\over7} (\ell_3 \cdot p_{12} - p_1 \cdot p_2) $ \\   \quad $ \times [\ell_5 \cdot (7p_1 - p_2) + \ell_6 \cdot (p_1 - 7 p_2)] \big\} $\end{tabular} 
 & $ (2\, N_c^4 + 24\, N_c^2) \, \delta_{a_1 a_2} $ & $4$ \cr \hline
\hline
(27) $\begin{matrix}  ~ \\ ~\\ ~\\ ~\\  ~ \\ ~ \\~\\~ \\~ \end{matrix}$   & 
\begin{tabular}{c} $  -(\ell_3\cdot p_1)^2 -(\ell_3\cdot p_2)^2 - 6(\ell_3\cdot p_1) (\ell_3\cdot p_2) $ \\ $ + (\ell_3\cdot p_1) ( \ell_4\cdot (p_1 + 5 p_2) ) $ \\ $ - (\ell_3 \cdot p_2) ( \ell_4 \cdot ( 3p_1 - p_2))  $ \\ $ + (p_1 \cdot p_2) \big[ 2 \ell_3 \cdot (p_{12} + \ell_3 - \ell_4)   $  \\ $ + \ell_4 \cdot (p_1 - p_2) - p_1 \cdot p_2 \big]   $ \\ $ + (\alpha_1 + 1) \big[ (\ell_3 \cdot p_{12} - p_1 \cdot p_2)^2 $  \\ $  - {1\over7} (\ell_3 \cdot p_{12} - p_1 \cdot p_2)(\ell_4 \cdot (7p_1 - p_2)) $ \\ $  - {2\over7} (\ell_3\cdot (\ell_3 - p_{12}) + p_1\cdot p_2)(p_1\cdot p_2) \big]  $ \end{tabular} 
& $  24 \,N_c^2 \, \delta_{a_1 a_2} $ & $1$ \cr \hline 
(28) $\begin{matrix} ~ \\ ~   \end{matrix}$   & $N_{27}$ & $ 24 \,N_c^2  \, \delta_{a_1 a_2} $ & $1$ \cr \hline
(29) $\begin{matrix} ~ \\ ~   \end{matrix}$   & $N_{27}$ & $ 24 \,N_c^2  \, \delta_{a_1 a_2} $ & $1$ \cr \hline
(30) $\begin{matrix} ~ \\ ~   \end{matrix}$   & $N_{27}$ & $ (2\, N_c^4 + 24\, N_c^2) \, \delta_{a_1 a_2} $ & $1$ \cr \hline
\hline
\end{tabular} 
\end{center}


\newpage 
\begin{center}
\begin{tabular}{c } 
{\bf Table 5 (continued)}. The result for the two-point four-loop form factor.
\end{tabular}
\end{center}

\begin{center}
%
\begin{tabular}{l | c | c | c} 
Graph &  Numerator factor  &  Color factor & \begin{tabular}{c}  Symmetry \\  factor \end{tabular} \cr \hline
\hline
(31) $\begin{matrix} ~ \\ ~ \\~ \\~\\~ \end{matrix}$   & 
\begin{tabular}{c} $ -4 (\ell_3\cdot p_1 - p_1\cdot p_2) (\ell_3\cdot p_2 - p_1\cdot p_2) $ \\ $ + (\alpha_1 + 1) \big\{ (\ell_3 \cdot p_{12} - p_1 \cdot p_2)^2 $ \\ $  - {2\over7} (\ell_3 \cdot \ell_3)(p_1 \cdot p_2) - {1\over7} (\ell_3 \cdot p_{12} - p_1 \cdot p_2) $ \\ $ \times [ \ell_3 \cdot (7p_1 - p_2) - 3 p_1 \cdot p_2 ] \big\} $ \end{tabular}
 & $ (2\, N_c^4 + 24\, N_c^2) \, \delta_{a_1 a_2} $ & $1$ \cr \hline
(32) $\begin{matrix} ~ \\ ~   \end{matrix}$   & $N_{31}$ & $  24 \,N_c^2 \, \delta_{a_1 a_2} $ & $1$ \cr \hline
(33) $\begin{matrix} ~ \\ ~   \end{matrix}$   & $N_{31}$ & $ 24 \,N_c^2  \, \delta_{a_1 a_2} $ & $1$ \cr \hline
(34) $\begin{matrix} ~ \\ ~   \end{matrix}$   & $N_{31}$ & $ 24 \,N_c^2  \, \delta_{a_1 a_2} $ & $1$ \cr \hline
\end{tabular} 
\end{center}


\section{Conclusion and outlook}

In this article the extension of ideas concerning color-kinematic duality from scattering amplitudes to form factors has been studied. The focus in this article has been the consequences of this extension at the loop level. In several examples it has been demonstrated that the duality leads to all known results in the literature up to three loops for stress-energy multiplet form factors in ${\cal N}=4$ SYM. While these results are obtained in a simpler and more uniform way than the original derivations, they also provide non-trivial support for the conjectured duality and the validity of the construction.

The same construction has also been applied at the four-loop level for the two-point form factor. A four-loop solution has been found that satisfies all color-kinematic relations and passes a considerable number of non-trivial unitarity  checks. An interesting feature in this approach is that planar and non-planar integrals are obtained simultaneously. The integration of the resulting set of integrals is work in progress \cite{inprogress}. The five-loop two-point integrand should also be obtainable with the same methods.

Our methods should also be applicable to form factors in less supersymmetric theories such as QCD. This is definitely a direction worth pursuing further as the duality relates in general planar and non-planar contributions. The interplay between these in the duality is in general a phenomenon which will be interesting to explore further. In the current form color-kinematic duality is only applicable to matter in the adjoint representation; a possible extension to matter in the fundamental representation would be very interesting as well. 

More generally, it should be fruitful to see how far and wide the general ideas of color-kinematic duality can be applied beyond scattering amplitude computations. These can be taken as a hint that there is sense to be made of the duality at a Lagrangian level in the gauge theory. Indeed, the existence of a Lagrangian formulation of gauge theory which yields an explicitly color-dual perturbation theory would be a natural explanation of the duality for form factors. This has so far only been achieved for low numbers of points \cite{Bern:2010yg} or in the self-dual sector of the gauge theory \cite{Monteiro:2011pc}. Moreover, it would be interesting to find the observable in the gravity theory which corresponds to the `square' of our form factor results.

\acknowledgments
It is a pleasure to thank Reinke Sven Isermann, Henrik Johansson, Congkao Wen, and in particular Andreas Brandhuber, Johannes Henn and Gabriele Travaglini for very useful discussions. RB would like to thank the Institute for Advanced Study for hospitality while this article was being finished. This work was supported by the German Science Foundation (DFG) within the Collaborative Research Center 676 ``Particles, Strings and the Early Universe''. The figures were generated using Jaxodraw \cite{jaxodraw1, jaxodraw2}, based on Axodraw \cite{Axodraw}.

\appendix
\section{Four-loop Jacobi relations \label{jacobi-4loop}}


In this appendix we give a set of Jacobi relations of the four-loop two-point form factor, from which one can generate all numerators by using the two master integral numerators: $N_{13}, N_{21}$.
\bea && N_2 = N_{10}[p_1, p_2, \ell_3, \ell_4, \ell_6, p_{12} - \ell_3 + \ell_4 - \ell_5]  + N_{10}[p_1, p_2, p_{12}- \ell_3,  -\ell_4 + \ell_5, \ell_6, \ell_3 - \ell_4]  \, , \\ 
 && N_1 = N_3 = N_4 = N_5 = N_2 \, , \\ 
 && N_6 = N_{13}[p_1, p_2, \ell_3, \ell_5, p_{12} - \ell_4, \ell_6] + N_{17}[p_1, p_2, \ell_3, \ell_3 - \ell_4 + \ell_5, \ell_5, \ell_6]  \, , \\ 
 && N_8 = - N_7 = N_6 \, , \\ 
 && N_9 = N_{13}[p_2, p_1, p_{12}- \ell_3, \ell_6, \ell_4,  \ell_5] + N_{19}[p_1, p_2, \ell_3, \ell_4, \ell_5, p_{12}-\ell_5- \ell_6]  \, , \\ 
 && N_{11} = - N_{10} = N_9  \, , \\
 && N_{12} = N_{10}[p_2, p_1, p_{12} - \ell_3, \ell_5 + \ell_6 - \ell_4, \ell_6, \ell_3 - \ell_4] - N_{17}[p_1, p_2, \ell_3, \ell_3 - \ell_4, p_{12} - \ell_5 - \ell_6, \ell_6] \, , \nonumber\\ &&\\
  && N_{16} = - N_{14} = - N_{15} = N_{13} \, , \\
   && N_{18} = N_{28}[p_1, p_2, \ell_3, \ell_4, p_1 - \ell_5, p_2 - \ell_6] - N_{28}[p_1, p_2, p_{12} - \ell_3, \ell_5 - \ell_4, p_1 - \ell_5, p_2 - \ell_6] \, , \nonumber \\ && \\ 
 && N_{17} = - N_{18} \, , \\
 && N_{20} = N_{28}[p_1, p_2, p_{12} - \ell_3, p_1-\ell_5, \ell_3 - \ell_4, \ell_6] - N_{28}[p_2, p_1, p_{12}-\ell_3, p_2 - \ell_6, \ell_4, \ell_5] \, , \\ 
 && N_{19} = N_{20} \, , \\
 && N_{22} = N_{23} = N_{24} = N_{25} = N_{21} \, , \\
 && N_{26} = N_{25} - N_{12} \, , \\
 && N_{28} = N_{13} + N_{21} \, , \\ 
 && N_{27} = N_{29} = N_{30} = N_{28} \, , \\
 && N_{31} = N_{13}[p_1, p_2, \ell_3, p_1 - \ell_5, \ell_4, p_2 - \ell_6] + N_{28}[p_1, p_2, \ell_3, \ell_3 + \ell_5, \ell_4, p_2-\ell_6] \, , \\ 
 && N_{32} = N_{33} = N_{34} = N_{31}  \, .
\eea
Note that $N_i = N_i[p_1, p_2, \ell_3, \ell_4, \ell_5, \ell_6]$ if not specified.


\section{Classifying color factors for two-point correlation functions to eight loops}\label{app:color}

In this appendix the independent group theory structures are classified which appear in the calculation of the $l$-loop two-point form factor in any gauge theory with adjoint matter only. Maximally supersymmetric Yang-Mills theory is a particular example of such a theory. The main observation is that these will always be proportional to $\delta^{ab}$. Contracting this with itself gives a vacuum graph in color space, times a universal factor which is the number of generators in the adjoint. In $SU(N)$ this number is $N_A = (N^2 -1)$. The classification in this appendix applies also to general two-point correlation functions for two operators in the adjoint representation. 

All vacuum graphs of a fixed number of loops consisting of only three-vertices can be generated using DiaGen \cite{DiaGen}, excluding self-energies and selecting 1PI graphs. In a non-supersymmetric theory the self-energy graphs are non-trivial, but will only contain color structures obtained at a lower-loop order, multiplied with a power of $C_A$. 

The output of the diagram generator is parsed for each graph by Mathematica into a definite color trace by assigning a structure constant to each three vertex. This index contraction is passed to a FORM \cite{Vermaseren:2000nd} program which itself uses the COLOR package \cite{vanRitbergen:1998pn}. After running the program the output of this is read back into Mathematica. This cycle is repeated for each graph in the output. 

\begin{table}
\begin{centering}
\begin{tabular}{c|c}\label{tab:loopcolor}
loops & color structures\\
\hline
1 	& 	$C_A$\\
2 	& 	$C_A^2$ \\
3 	& 	$C_A^3$ \\
4 	& 	$C_A^4$ \quad $\tilde{d}_{44}$ \\
5 	& 	$C_A^5$ \quad $\tilde{d}_{44} C_A $ \\
6 	& 	$C_A^6$ \quad $\tilde{d}_{44} C_A^2 $ 	\quad $\tilde{d}_{444}$ \\
7 	& 	$C_A^7$ \quad $\tilde{d}_{44} C_A^3 $ 	\quad $\tilde{d}_{444} C_A$ 	\quad $\tilde{d}_{644}$       \\[3pt]
8 	& 	$\left\{\begin{array}{c} C_A^8 \quad \tilde{d}_{44} C_A^4   	\quad \tilde{d}_{444} C^2_A 	\quad \tilde{d}_{644} C_A  \\
 		\tilde{d}_{664} \quad   \tilde{d}_{844}  \quad  \tilde{d}_{4444a}  \quad \tilde{d}_{4444b} \quad N_A \tilde{d}_{44}^2 \end{array}\right\}$
\end{tabular}
\caption{Classification of group theory invariants in a two-point calculation at a fixed number of loops. Tildes indicate that the `d'-type invariants in the output of COLOR have to be divided by a common factor of $N_A$}
\end{centering}
\end{table}

The result is a list of group theory factors for each separate graph, given in table 6. The labels for the invariants are produced by the COLOR package. By computing a small choice of color traces by explicit Fierzing (which is many orders of magnitude slower than the COLOR package) we obtain up to seven-loop order
\begin{align}
C_A & = N \, , \\
N_A & =  N^2 -1\, , \\
\tilde{d}_{44} & =  \frac{1}{24}  \left(36 \, N^2+ N^4\right) \, ,\\ 
\tilde{d}_{444} & = \frac{1}{216} \left(324 \,N^2  + 135\, N^4 + N^6)\right) \, , \\
\tilde{d}_{644} & = \frac{1}{1920} \left(3564 \,N^3 + 225\, N^5 + N^7 \right)\, .
\end{align}
The form factor at $l$ loops is proportional to $g_{\textrm{ym}}^{2l}$, which is usually combined with $N$ into the 't Hooft coupling $\lambda \equiv N g_{\textrm{ym}}^{2}$. In terms of $\lambda$ and $N$ it is seen from table 6 that the first non-planar correction sets in at four loops, while the next-to-non-planar correction starts at six loops. This pattern is expected to continue based on the general arguments of \cite{'tHooft:1973jz}. Note that these results can be used to reverse-engineer the color factor of any result given for $SU(N)$ two-point functions to more general invariants.

The number of different color structures appearing at a certain loop order for two-point functions comes down to classifying trivalent graphs at this loop order with two endpoints, modulo Jacobi relations. In particular, the two external legs can be counted as identical. This turns out to be a mathematics problem which has been studied in various places, see \cite{Broadhurst:1997pe} and references therein. This paper provides a conjecture for the number  $c_l$ of  independent Casimirs appearing at loop order $l$ as a generating function:
\begin{align}
\sum_{l=0}^{\infty} c_{l+1} y^l & = \frac{y^2 -1  - y^5}{(y-1)^3 (y+1) (y^2 + y + 1)} \\
& = 1 + y + y^2 + 2 y^3 + 2 y^4 + 3 y^5 + 4 y^6 + 5 y^7 + 6 y^8 + 8 y^9 + 
 9 y^{10} + 11 y^{11} + \mathcal{O}\left( y^{12}\right)
\end{align}
Our explicit computations are in agreement with these numbers up to seven loops. The discrepancy at eight loops is most likely resolved by finding relations between the Casimir invariants COLOR introduces at this loop order: this set is not guaranteed to be minimal beyond seven loops by the used methods.

\bibliographystyle{jhep}

\bibliography{ff}

\end{document}